\documentclass{aa}%
\usepackage{longtable}
\usepackage{amsmath}
\usepackage{graphicx}
\usepackage{natbib}
\usepackage[varg]{txfonts}
\usepackage{savesym}
\usepackage{color}

\newcommand{\XMM}{\emph{XMM-Newton}}
\newcommand{\Swift}{\emph{Swift}}
\newcommand{\ergx}{\,erg\,s$^{-1}$\,cm$^{-2}$}
\newcommand{\ergs}{\,erg\,s$^{-1}$}

\newcommand{\XSPEC}{{\sc XSpec}}

\begin{document}

\title{\XMM\ observations of eleven intermediate polars and possible candidates}

\author{
H. Worpel\inst{1}\fnmsep\thanks{Corresponding author:{hworpel@aip.de}},
A.~D. Schwope\inst{1},
I. Traulsen\inst{1},
K. Mukai\inst{2,3},
S. Ok\inst{4,1}
}
\institute{Leibniz-Institut f\"ur Astrophysik Potsdam (AIP), An der Sternwarte 16, 14482 Potsdam, Germany \and
           CRESST and X-ray Astrophysics Laboratory, NASA/GSFC, Greenbelt, MD 20771, USA\and
           Department of Physics, University of Maryland, Baltimore County, 1000 Hilltop Circle, Baltimore, MD 21250, USA
\and Department of Astronomy \& Space Sciences, Faculty of Science, University of Ege, 35100, Bornova, \.Izmir, Turkey
}
\authorrunning{Worpel et al}
\date{}


\abstract
{}
{We aim to identify new intermediate polars (IPs) in \XMM\ observations from a list of promising candidates. By selecting targets not previously known to be X-ray bright we aim to uncover evidence for an X-ray underluminous IP subpopulation.}
{We performed period searches on the \XMM\ X-ray and optical data of our targets to seek both the spin and orbital periods, which differ in IPs. We also investigated the X-ray spectra to find the hot plasma emission shown by these objects. With archival \Swift\ data we coarsely investigated the long-term X-ray variability, and with archival optical data from a variety of catalogues, we compared the optical to X-ray luminosity to identify X-ray faint objects. This paper presents the first \XMM\ observation of the prototype IP, DQ Her.}
{We find firm evidence for HZ~Pup, V349~Aqr, and IGR~J18151-1052 being IPs, with likely white dwarf spin periods of 1552, 390, and 390\,s, respectively. The former two have luminosities typical of IPs, and the latter is strongly absorbed and with unknown distance. GI~Mon and V1084~Her are apparently non-magnetic CVs with interesting short-term variability unrelated to WD spin. V533~Her is probably a magnetic CV and remains a good IP candidate, while V1039~Cen is possibly a polar. The remaining candidates were too faint to allow for any firm conclusions.}
{}

\maketitle

\section{Introduction}

Intermediate polars (IPs) are cataclysmic variables (CVs) in which the
white dwarf's (WD) strong magnetic field ($\sim 10$~MG) suppresses the
inner part of the accretion disc and largely governs the flow of
accreting matter, but it is not powerful enough to synchronise the WD's
spin to the binary orbit. Material reaches the WD surface by coupling onto
magnetic field lines and falling quasi-radially. IPs are therefore important laboratories for studying accretion processes in a variety of modes and in an environment of strong magnetic and gravitational fields. For a review, see , for example, \cite{Patterson1994}.

The radially infalling matter forms a shock above the WD surface
at the magnetic pole. Copious X-rays are emitted in the post-shock region in
the form of bremsstrahlung, with luminosity typically reaching $\sim 10^{33}$~erg/s,
well-described by a plasma temperature of 20-50~keV, including prominent
iron emission lines at 6.4-7.0~keV. These spectra, depending on the shape of the
accretion column, can appear as either cooling flows or as photoionisation spectra (for example 
\citealt{MukaiEtAl2003}). The amount of intrinsic absorption is highly variable among systems
and, although most IPs exhibit a suppression of X-ray photons at low ($<2$~keV) energies, \emph{ROSAT} and \emph{XMM-Newton} have uncovered 'soft IPs', with low intrinsic absorption and a pronounced black-body like component at low energies ($kT_{\rm bb} \simeq 50$\,eV, for example,  
\citealt{MasonEtAl1992, StaudeEtAl2006}). The magnetic axis is usually offset from the WD spin
axis, leading to modulations of the optical and X-ray light curves at the spin period.

Possible IP candidates are usually discovered in optical surveys such as
SDSS or IPHAS on the basis of their distinctive emission lines, or in X-ray
surveys from their hard X-ray spectra. Confirming a candidate depends on
several observational signatures (for example \citealt{Patterson1994}). The most important
of these is the confirmation of two periods in both
X-ray and optical photometry. These demonstrate that the orbital and WD spin
periods are different, thus distinguishing them from the more strongly
magnetic, spin period synchronised, polars. The optical emission lines, and
a hard X-ray spectrum, indicate a magnetic CV. X-ray observations
with high timing and spectral sensitivity are therefore crucial for
identifying IPs; see, for example, \cite{NucitaEtAl2019} for a recent IP confirmation by this method.

IPs are also suspected to contribute much of the Galactic Ridge
X-ray Emission (GRXE), a diffuse-looking X-ray background concentrated
around the Galactic plane \citep{WorrallEtAl1982}, whose nature remained
unknown for over 20 years. The discovery of the 6.4-7.0~keV iron triplet
\citep{EbisawaEtAl2008} implies that its origin is plasma of temperature 2-10~keV.
A recent \emph{Chandra} observation determined that 80\% of the GRXE can be resolved into
numerous individually faint point sources \citep{RevnivtsevEtAl2009}. The nature of
these objects is still uncertain. Above 10~keV the IPs are expected to dominate
\citep{WarwickEtAl2014}, and at softer energies they are joined
by polars, dwarf novae, and coronally active main sequence stars. The IPs therefore seem 
to represent an important part of the X-ray emission of the Galaxy, and of other galaxies.

Even though there are likely many IPs in the Galaxy, only around fifty have been definitively identified as such. The X-ray underluminous subpopulation \citep{PretoriusMukai2014} is even more elusive, with only AE~Aqr, DQ~Her, and V902 Mon having previously been discovered in soft X-rays (for example \citealt{AungwerojwitEtAl2012}) and DO~Dra, V1025~Cen, and EX~Hya found to be underluminous in the Swift-BAT energy range of 14-195\,keV \citep{PretoriusMukai2014}. Furthermore, some of these may only appear faint because of obscuration by their accretion discs.

We reviewed the literature regarding all unconfirmed IP candidates, as listed in the \emph{iphome} list\footnote{\url{https://asd.gsfc.nasa.gov/Koji.Mukai/iphome/catalog/alpha.html}}. Typically, these targets have shown either two optical photometric periods or hard X-ray emission detected serendipitously, and a few have been classified as possible magnetic CVs on the basis of their optical emission lines. Some are known to be CVs solely because of a past nova eruption. From this list of candidates we excluded Galactic centre sources, which are likely to give very low X-ray flux and suffer from source confusion. Several other good candidates were not considered, because they did not have good observing opportunities with \XMM.

\emph{XMM-Newton} observing time was obtained for eleven IPs and IP candidates. Here we present the results, refine the X-ray spectral characteristics of this population, their X-ray to UV flux ratios, and seek spin period modulations. We use an observation of the prototype X-ray underluminous IP, DQ Her to develop the methods necessary to do this.

We also use archival \emph{Swift} data to characterise their long-term X-ray variability and utilise archival optical data, such as that collected by the AAVSO, to determine their long and short term optical variability. A twelfth IP candidate in this observing campaign, V902 Mon, was published already in \cite{WorpelEtAl2018}, in which we found that V902 Mon is an X-ray underluminous IP. A more extensive catalogue, based on the methodology of this paper, incorporating all archival and previously unpublished \XMM\ and \Swift\ observations of IPs and IP candidates will be presented in a subsequent paper.

A brief literature review for all sources considered in this campaign is given below in order of increasing right ascension.

\section{Review of sources}

\subsection{ GI Mon }

This object was confirmed as a CV through optical spectroscopy, but it has only a very weak H$\alpha$ line and the other emission lines were not detectable \citep{LiuHu2000}.
It probably has an orbital period of 4.33~hr and a tentative second period of 48.6~min, which is possibly the white dwarf spin period \citep{WoudtEtAl2004}. Longer photometry to confirm the shorter period would be a benefit. A 0.1 magnitude dip lasting 45 minutes may be an eclipse, but if so, the orbital period would have to be longer than 4.8 hours \citep{Rodriguez-GilTorres2005}. The system went nova in 1918 \citep{AdamsJoy1918}.

\subsection{HZ Pup}

The orbital period is 5.11~hr, and some periodicities around 1200-1400~seconds have been interpreted as the spin period and various beat periods \citep{AbbottShafter1997}, but these, so far, have been far from conclusive. Its distance is estimated to be about 1~kpc \citep{PretoriusKnigge2008}. This system went nova in 1963 \citep{Hoffmeister1964}.

\subsection{V597 Pup}
This CV system underwent a nova eruption in 2007 \citep{PereiraEtAl2007}. It became visible in soft X-rays in January 2008 \citep{NessEtAl2008}, and later photometry showed it to be eclipsing with a period of 2.67 hr and a possible spin period of 524\,s \citep{WarnerWoudt2009}.

\subsection{V1039 Cen}

This CV has an orbital period of 5.92 hr, and a subtle second period of 720~s interpreted as possibly the spin period \citep{WoudtEtAl2005}. 
The system underwent a nova eruption in 2001 \citep{LillerEtAl2001}. 

\subsection{V1084 Her (1RXS J164345.2+340236)}

This source is an optically bright CV with a 2.6~hr or 2.9~hr orbital period \citep{MickaelianEtAl2002}, thought to be an SW~Sex star \citep{PattersonEtAl2002}. \cite{RodriguezGilEtAl2009} found a periodicity of 19.4\,min which they interpreted as half of the beat period between the WD spin and orbital periods. This implies a spin of 31.7\,min (for the longer possible orbital period) but there are optical quasi-periodic oscillations (QPOs) of $\sim$ 1000~s so the spin period is still uncertain. It is a somewhat faint source in the ROSAT All-Sky Survey (RASS) at 0.05 c/s \citep{VogesEtAl1999}.

\subsection{DQ Her}
DQ Her is the prototype star for the intermediate polars and the subject of a vast body of academic literature too extensive to summarise here; see \cite{Patterson1994} for a
review. It is a deeply eclipsing system with a probable 71~s spin period \citep{Walker1956}, though \cite{ZhangEtAl1995} show strong evidence that the spin period is actually
twice as long, and that its orbital period is 4.65 hours. Its distance was estimated to be about 525~pc by \cite{VaytetEtAl2007}, but Gaia data shows it to be somewhat closer ($494\pm6$\,pc; see Table \ref{tab:Gaia}). The eclipse timings show a 14~yr sinusoidal modulation, but this is unlikely to be due to a circumbinary third object \citep{PattersonEtAl1978}. This system went nova in 1934 \citep{Stratton1934}.

DQ Her is quite X-ray faint. It was not detectable in \emph{Swift-BAT} \citep{BrunschweigerEtAl2009}. A series of pointed ROSAT observations totalling 13\,ks gave a flux of $1.2\times 10^{-13}$~erg~s$^{-1}$~cm$^{-2}$ in the 0.1-2~keV band, with no evidence of X-ray eclipses \citep{SilberEtAl1996}. It was fainter by a factor of 3 in two \emph{Chandra} observations, totalling 69\,ks, and showed shallow X-ray eclipses \citep{MukaiEtAl2003}.

\subsection{IGR J18151-1052}

This object is a moderately bright, hard X-ray source discovered by INTEGRAL (0.018 c/s in the Swift 1-10~keV range, \citealt{KrivonosEtAl2009}). 
\citet{MasettiEtAl2013} presented optical spectra of INTEGRAL discovered X-ray sources and classify this object as a CV, likely a magnetic CV, based on the appearance of HeI and HeII lines. The spectrum appeared reddened through strong absorption in the interstellar medium, perhaps not too surprising given its location at only two degrees above the galactic plane (galactic latitude 18 degrees). Its X-ray flux in Swift-BAT was $1.4\times 10^{-11}$~erg~s~$^{-1}$~cm$^{-2}$ in the 14-195 keV band \citep{CusumanoEtAl2010}. Its possible magnetic CV nature and hard X-ray spectrum make it an IP candidate, though as yet there is no information regarding its possible periods.

\subsection{ V4745 Sgr }

Nova Sgr 2003 \citep{BrownEtAl2003} was suggested as an IP candidate on the basis of a 5 hour orbital period and a possible spin of 1489~s \citep{DobrotkaEtAl2006}.
Its distance was roughly estimated to be between 9 and 19~kpc \citep{CsakEtAl2005}, but Gaia measurements suggest it is significantly closer; see Table \ref{tab:Gaia}. It was detected in a \emph{Swift} observation \citep{SchwarzEtAl2011}.

\subsection{V533 Her}
This system underwent a nova eruption in 1963. It has an orbital period of 3.53~hr, and phase-dependent absorption in some of its emission lines suggest that it is an SW Sextantis star \citep{ThorstensenTaylor2000}. A 63.6~s pulsation observed by \cite{Patterson1979} is very likely to be the spin period, but it had vanished within a few years (for example \citealt{RobinsonNather1983}) and has not been seen since. Another period, 23.33 min, is suggested \citep{Rodriguez-GilMartinez-Pais2002} based on variations in the equivalent width of emission lines.

\subsection{ V1425 Aql (Nova Aql 1995) }
This CV went nova in 1995 \citep{NakanoEtAl1995}, and was found to have resumed accretion just 15 months later by \cite{RetterEtAl1998}. Those authors found three periodicities: 6.14 hours, 5188~s, and 6825~s, which they interpreted as the orbital, spin, and beat periods, respectively.

\subsection{SDSS J223843.84+010820.7 (V349 Aqr)}

Recognized as a CV by \cite{BergEtAl1992}, it is a probable intermediate polar. A 6.7~min spin period is likely and its orbital period is either two hours or, more likely, 3.226 hours \citep{SzkodyEtAl2003, WoudtEtAl2004, SouthworthEtAl2008}.

\section{Observations and Methods}

\subsection{XMM-Newton}

The core of our observing campaign is eleven \XMM\ observations that were performed as part of AO17-080411. These were taken between 2017 Apr and 2019 Mar, giving a total of 388.5\,ks of observing time. All were performed in the same observing modes to ensure consistency between observations.

\subsubsection{EPIC and RGS}

\XMM\ carries five X-ray instruments: the EPIC-$pn$ camera \citep{StruderEtAl2001}, two EPIC-MOS \citep{TurnerEtAl2001} instruments, and two Reflection Grating Spectrometers (RGS; \citealt{denHerderEtAl2001}). Owing to the faintness of the sources, we have not used the RGS in this study.

For each source we extract EPIC spectra with a minimum of 16 counts per bin, and analyse these with version 12.9.1n of \texttt{Xspec} \citep{Arnaud1996}. The choice of 16 counts per bin is a balance between retaining the validity of the $\chi$ statistic and retaining fine spectral resolution. Counts above about 15 have generally been considered sufficiently Gaussian \citep{HumphreyEtAl2009}. Although the C-statistic \citep{Cash1979} is believed to be less biased than $\chi^2$ and can be used with few counts per spectral bin obtaining uncertainties on fit parameters with it is laborious \citep{Kaastra2017} and the correct methods have not yet been incorporated into spectral fitting tools. We therefore stay with $\chi^2$ and reserve the use of C-stat for fine spectral analyses with few counts per bin. 

Unless otherwise specified, EPIC observations were performed in full frame mode with the thin filter and the OM operated in fast mode with the UVW1 filter. The X-ray data were reduced with the {\tt emchain} and {\tt epchain} tasks for the MOS and $pn$ cameras respectively, and the OM data were reduced with the {\tt omichain} and {\tt omfchain} tasks. The sizes of the source and background extraction regions were different from object to object, and were chosen to maximise signal-to-noise.

For some spectra where fine features such as emission lines are sought we also analysed unbinned spectra and obtain uncertainties on the best fits using the methods developed by \cite{Kaastra2017}. \XMM\ spectra were fit for all three EPIC instruments jointly, between 0.2 and 10.0\,keV. For the determination of bolometric X-ray fluxes we then extended the energy range using \XSPEC's \texttt{energies} command and determined the flux with the \texttt{cflux} convolution model with all other parameters temporarily frozen. Uncertainties on spectral parameters and flux determinations were found using the \texttt{steppar} command and are at the $1\sigma$ level. We assumed Solar abundances, using the abundance table of \cite{AndersGrevesse1989} unless otherwise specified. Since galactic $n_H$ along the line of sight to our sources is an order of magnitude smaller than $n_H$ intrinsic to the source (see Table \ref{tab:Gaia}), we did not include it in our spectral fits.

We also gave bolometric luminosities, converted from the fluxes via the distance determinations and uncertainties on them, for each of the sources given in the catalogue of \cite{Bailer-JonesEtAl2018}. Uncertainties on luminosity were calculated with the standard propagation of errors, assuming the uncertainties on the fluxes and distances are two-sided normal distributions.

The photon arrival times were corrected to the Solar System barycentre using the \texttt{barycorr} task. We used the H-test \citep{deJagerEtAl1989} to search for periodicities in the photon arrival times. This test resembles the Rayleigh method but is more powerful for detecting pulses of arbitrary shape, including multiple-peaked profiles. A spinning WD with two accreting poles may show such a two-peaked profile.

\subsubsection{Optical monitor}

The optical monitor (OM; \citealt{MasonEtAl2001}) on \XMM\ is an optical/ultraviolet telescope with a diameter of 30\,cm. It provides  time-resolved optical data simultaneous with the X-ray observations. We operated this telescope with the UVW1 filter in imaging and fast modes in all observations. The UVW1 filter has an effective wavelength of 2910\AA\ \citep{KirschEtAl2004}.

For each observation we barycentre corrected the light curve time stamps and performed a period search using the Analysis-of-Variance (AoV) method (for example \citealt{Schwarzenberg-Czerny1989}) to attempt to find an IP spin period.

\subsection{Swift}
\label{sec:swift}
The X-Ray Telescope (XRT; \citealt{BurrowsEtAl2005}) on \Swift\ is a focusing instrument sensitive in the 0.3-10\,keV range. We sought all \Swift\ observations of intermediate polar candidates within the 11.8 arcmin radius of \Swift\ XRT's field of view.

We found that many of these observations of IP candidates were short duration or serendipitous off-axis exposures. Thus, when a source is visible in the Swift XRT data; see Section \ref{sec:swiftxrt}, there will typically not be enough photons to extract a spectrum. To study the long-term X-ray luminosity of the source we therefore proceed as follows. We assumed that the source exhibits the same spectral shape for the \emph{Swift} observations as it does for the \XMM\ pointing, with possibly different levels of normalisation.

We then calculated background subtracted counts and used the \texttt{fakeit} of Xspec to calculate the spectrum normalisation necessary to deliver that number of photons. Since our source and background counts are low, Poissonian statistics apply and it is not correct to assume $X$, the subtracted counts, are equal to the source region counts $s$ minus the area weighted background counts $b/A$. Instead, the probability distribution function for $X$ is:

\begin{equation}
    P(X|s,b,A) \propto e^{-X}\displaystyle\sum_{i=0}^{s}\binom{s}{i}(b+s-i)![(A+1)X]^i,
    \label{eq:loredo}
\end{equation}
as shown by \cite{Loredo1992}. We obtained $1\sigma$ upper and lower uncertainties using the procedure described in \cite{KraftEtAl1991}, which yields the smallest interval containing both the peak of the distribution and 68.27\% of its area. The luminosity derived for the \XMM\ observation then gives an estimate for the luminosity of the \Swift\ observations, assuming the same spectral shape.

\subsubsection{X-ray Telescope}
\label{sec:swiftxrt}
We reduced the XRT data using the \Swift\ {\sc xrtpipeline} task and corrected the photon arrival times to the Solar System barycentre using the {\sc barycorr} task of FTOOLS \citep{Blackburn1995}.

\subsubsection{Burst Alert Telescope}

\Swift\ also carries the high energy coded aperture Burst Alert Telescope, but our sources are too faint to be detected in the BAT.

\subsubsection{Ultraviolet Optical Telescope (UVOT)}

The UVOT instrument is a visual/ultraviolet telescope analogous to the OM on \XMM. Although the exposure times were generally too short to perform period searches for all but a few of the longest observations, we used the UVOT to detect general optical variability and populate spectral energy distributions (SEDs).

\subsection{Optical observations}

\subsubsection{AAVSO}

We downloaded all AAVSO observations of the targets, in all available energy bands.  We discarded any that were limiting magnitudes, that is, nondetections. The timestamps were corrected from JD to the Solar System barycentre using light travel times with Astropy \citep{Astropy2013, Astropy2018}. Since the observations were performed at numerous locations on Earth, but light travel time from one point on Earth to another is negligible compared to the uncertainties in the time stamps, we used the centre of the Earth as the geographic reference point.

With the barycentre-corrected time stamps we also performed AoV period searches, thus making our analysis sensitive to periodicities present over timescales longer than the \XMM\ pointings.

Several sources also have white-light photometry from the Catalina Sky Survey (CSS; \citealt{DrakeEtAl2009}). For these we add the barycentre-corrected CSS points to the AAVSO points to increase the sensitivity of our period searches.

\subsubsection{TUBITAK}

We performed one observation of GI~Mon with the 1\,m T100 telescope at TUBITAK National Observatory on the night of 2018 Jan 26. The data were reduced with IRAF \citep{Tody1993}. The exposures were 70\,s long and performed with the V filter.

\subsubsection{Other archives}

To aid in characterising the X-ray to optical luminosities of the IP candidates we constructed an SED for each source. The SEDs are populated with data points from \XMM and Swift X-ray data points where available.

The low energy segment of the SEDs incorporates data from the optical telescopes on XMM and Swift, as well as data from AAVSO. We have also used archival data from surveys such as the Sloan Digital Sky Survey (SDSS; \citealt{BlantonEtAl2017}), the Two-Micron All Sky Survey (2MASS; \citealt{SkrutskieEtAl2006}), the Wide-field Infrared Survey (WISE; \citealt{WrightEtAl2010}), the Galaxy Evolution Explorer (GALEX \citealt{MartinEtAl2005}), the Panoramic Survey Telescope and Rapid Response System (Pan-STARRS; \citealt{ChambersEtAl2016}), the Visible and Infrared Survey Telescope for Astronomy (VISTA; \citealt{CrossEtAl2012}), and SkyMapper \citep{WolfEtAl2018}.

For each source we plotted the frequency-dependent luminosity ($\nu L_\nu$), and the uncertainties incorporate both the uncertainties on the flux measurements and on the distance to the source. For V1425~Aql and IGR~J18151$-$1052, where we have no distance estimate, we plotted the frequency-dependent flux ($\nu F_\nu$) instead. The effective wavelengths and zero points for each instrumental filter are given in table \ref{tab:filters}.

\subsubsection{Gaia distances}

The Gaia telescope has, in its second data release, provided parallax measurements for most of the IP candidates in our sample \citep{Gaia2018, Gaia2016}. Only V1425~Aql and IGR~J18151$-$1052 could not be identified in the Gaia catalogue. Distances, and uncertainties on the distances, have been calculated via a probabilistic model by \cite{Bailer-JonesEtAl2018}. The distances and their $1\sigma$ uncertainties are given in Table \ref{tab:Gaia}.

\begin{table}
\caption{Distances to IP candidates in pc from the Gaia catalogue. No distance measurement could be obtained for V1425 Aql or for IGR J18151-1052. Galactic $n_H$ are derived from the 3D dust maps of \cite{BekhtiEtAl2016}. Sources with unknown distances or declinations south of $-30^\circ$ are omitted.}
\begin{tabular}{lll}
Source & Distance (pc) & $n_H$ (cm$^{-2}$)\\\vspace{2pt}
GI Mon& $2850^{+500}_{-400}$&$4.0\times 10^{20}$\\\vspace{2pt}
HZ Pup& $2300^{+400}_{-300}$&$4.0\times 10^{20}$\\\vspace{2pt}
V597 Pup& $3000^{+2200}_{-1300}$&\textellipsis \\\vspace{2pt}
V1039 Cen& $2200^{+2270}_{-890}$&\textellipsis \\\vspace{2pt}
V1084 Her& $434\pm 6$&$1.0\times 10^{20}$\\\vspace{2pt}\\\vspace{2pt}
DQ Her& $494\pm 6$&$2.5\times 10^{20}$\\ \vspace{2pt}
IGR J18151-1052 & \textellipsis&\textellipsis \\\vspace{2pt}
V4745 Sgr& $4600^{+3800}_{-2000}$&\textellipsis \\\vspace{2pt}
V533 Her& $1165^{+45}_{-42}$&$3.0\times 10^{20}$\\\vspace{2pt}
V1425 Aql& \textellipsis&\textellipsis \\\vspace{2pt}
V349 Aqr& $2200^{+900}_{-600}$&$2.5\times 10^{20}$\\\vspace{2pt}
\end{tabular}
\label{tab:Gaia}
\end{table}

\section{Results}

We report the results of each source in order of increasing right ascension, except that we put DQ~Her first as the prototype faint IP. The complete log of X-ray observations is given in Table \ref{tab:obslog}.

\subsection{DQ Her}

We observed DQ~Her for 41.9\,ks with \XMM\ on 2017 Apr 19. Our source extraction regions were circles of radius 20, 8.8, and 8.9 arcsec for the EPIC-$pn$, MOS1, and MOS2 cameras, respectively, and the background regions were large nearby circles. The observation was affected by moderately high background for the final 40\% of the exposure, with good time intervals (GTIs) judged by eye. Background-subtracted light curves were produced using the {\tt epiclccorr} task for all three EPIC cameras and are shown in the top panel of Figure \ref{fig:dqher_fullpage}. They are divided into time bins of 600\,s, and span an energy range of 0.2 to 10.0\,keV. The OM light curve has a 10\,s binning and is shown in the second panel of Figure \ref{fig:dqher_fullpage}.

The left central panels of the Figure show the X-ray and UV (top and bottom) light curves folded on the linear orbital ephemeris given in \cite{ZhangEtAl1995}. The X-ray data is divided into 20 bins and the UV data into 50 bins. The right middle panels show the data folded on the 71.067\,s spin period, with the eclipse ($0.875<\phi_\text{orb}<1.125$) omitted. 

The X-ray light curve shows clear variability, including a possible shallow eclipse feature in the second orbital cycle. It is, however, not apparent in the first cycle. The eclipse is obvious in the UV data, and is marginally detected in EPIC-$pn$. There is no hint of the spin period in either X-rays or UV. To roughly quantify any potential variability in the phase-folded light curves, we attempted to fit these to a constant. For the orbital phase folded data the probability of no variability was less than $10^{-3}$ for all three instruments. For the spin period we could not rule out the null hypothesis.

We applied our period search methods to DQ Her because it is eclipsing, and has a short spin period with a strong integer multiple alias. We combined the barycentre-corrected AAVSO and CSS data, as discussed above, and sought periodicities between 60\,s and 18,000\,s. The orbital period of 4.65 hours was obvious but we did not recover the likely spin period of 71 or 142\,s, which is unsurprising given the long exposure times of the optical observations. The H-test applied to the photon arrival times did not show the spin period, and the \XMM\ observation was too short to find the orbital period via the usual period searches. Since X-ray periodicities often manifest only at lower energies \citep{NortonWatson1989}, we divided the photon list in half by energy and ran the H-test again on the low and high energy halves individually. The division was at 0.99\,keV, and we did not find any signal.

Using the source and background regions described above, we extracted spectra for DQ~Her for all three EPIC instruments. We binned the spectra into groups and used {\tt Xspec} \citep{Arnaud1996} fit the data between 0.2 and 10.0\,keV for the MOS instruments and between 0.2 and 5.0\,keV for EPIC-$pn$ owing to very low counts at high energy with that instrument. None of the other sources were affected in this way so for all other targets we use the energy range 0.2--10.0\,keV for all instruments. We obtained a good spectral fit to the X-ray data with a partially covered two-temperature APEC plasma. The fit is summarised in Table \ref{tab:dqherspec} and displayed in the bottom left panel of Figure \ref{fig:dqher_fullpage}. The plasma temperatures are rather low, suggesting extensive reprocessing.

\begin{table}
\caption{Spectral fit for DQ Her. Absorbed fluxes and luminosities are calculated in the 0.2-10.0\,keV energy range, errors are 1$\sigma$ uncertainties}
\begin{tabular}{ll}
nH & $1.2\pm 0.3\times 10^{22}$\,cm$^{-2}$ \\
cvr. fract & $86^{+2}_{-3}\%$ \\
kT$_1$ & $0.27\pm 0.02$\,keV  \\
norm$_1$ & $9.6^{+2.4}_{-1.9}\times 10^{-5}$  \\ 
kT$_2$ & $0.84^{+0.6}_{-0.5}$\,keV \\
norm$_2$ & $7.0^{+1.6}_{-1.4}\times 10^{-5}$ \\
$\chi^2_\nu$ & 1.05 (151) \\ \hline
Flux   & $ 8.56\pm0.21\times 10^{-14}$\,erg\,cm$^{-2}$\,s$^{-1}$ \\
Lum    & $ 2.50\pm0.08\times 10^{30}$\,erg\,s$^{-1}$ \\
\end{tabular}
\label{tab:dqherspec}
\end{table}

We also saw strong emission from the iron line complex around 6.4-7.0\,keV. There were not enough photons above 5\,keV to reliably measure the continuum, so we cannot confidently measure the equivalent widths of the lines. However, the contribution of the 6.4\,keV fluorescent iron line is likely to be quite large because the best-fit line energy is artificially low. This is evidence of significant scattering in the DQ~Her system.

In the bottom right hand panel of Figure \ref{fig:dqher_fullpage} we show the SED of DQ~Her from infrared to X-rays. This plot has been assembled from numerous archives. The X-ray data have been binned by a factor of 50 to avoid cluttering the graph; we have done the same for the SEDs of all the other targets. It shows a clear stellar energy distribution up to about $10^{15}$\,Hz, and moderate variability in optical and UV-- as expected for a CV. The X-ray region again demonstrates the very soft spectrum. 

In the SED panel we indicate the relative flux uncertainty arising from the uncertainty in the Gaia distance measurement with a red marker. Since the distance to DQ~Her is known very precisely this marker is very small but for other sources, where we repeat the same procedure, it can be quite large. Measurement uncertainties are indicated on individual data points, although these are often also smaller than the marker used to plot them. In all cases uncertainties are at the $1\sigma$ level.

\subsection{GI Mon}
\label{sec:gimon}
Our \XMM\ observation of GI~Mon was 35\,ks long, and was not at all affected by soft proton flaring. Our source extraction regions were circles of radius 13.4, 8.8, and 11.3 arcsec for the EPIC-PN, MOS1, and MOS2 instruments, respectively. As for DQ~Her, we used large circular regions for the background. The background-subtracted light curves for the EPIC instruments and OM are shown in the top two panels of Figure \ref{fig:gimon_fullpage}. The X-ray and UV light curves display some variability but to the eye there is no obvious periodicity.

We performed a period search on the X-ray photon arrival times using the H-test method, but there was no significant signal. Similarly, the AoV method applied to the OM data showed no periodicities. In the central panels of Figure \ref{fig:gimon_fullpage} we show the X-ray and UV light curves folded on the suggested orbital and spin periods of \cite{WoudtEtAl2004}. There is no conspicuous modulation, but there may be a low-amplitude and irregularly-shaped signal at the orbital period. As for DQ Her, we attempted to fit this with a constant and found the hypothesis of no variability is disfavoured with a probability of less than $10^{-6}$. 

To estimate how intense a signal would have been detectable, we generated faked X-ray photon event lists with the same length and average count rate as the real observations, a period of 1,000 seconds, and a range of pulsed fractions. We then ran the H-test on these simulated lists. A pulsed fraction of 40\% would still have been detectable for GI~Mon. Similarly, we found that a 5\% pulsed fraction for the UV data would still have been detectable.

The X-ray spectrum of GI Mon could be adequately fitted with an absorbed one-temperature APEC model ({\tt phabs*apec}; spectral parameters and $\chi^2$ given in Table \ref{tab:gimonspec}). The luminosity in the 0.2-10\,keV range was around $4.7^{+1.2}_{-1.0}\times 10^{31}$erg\,s$^{-1}$ suggesting that, if it is an IP, it is around an order of magnitude fainter than a typical member of the class. The apparently high plasma temperature is also suggestive of this interpretation.

\begin{figure}
 \includegraphics{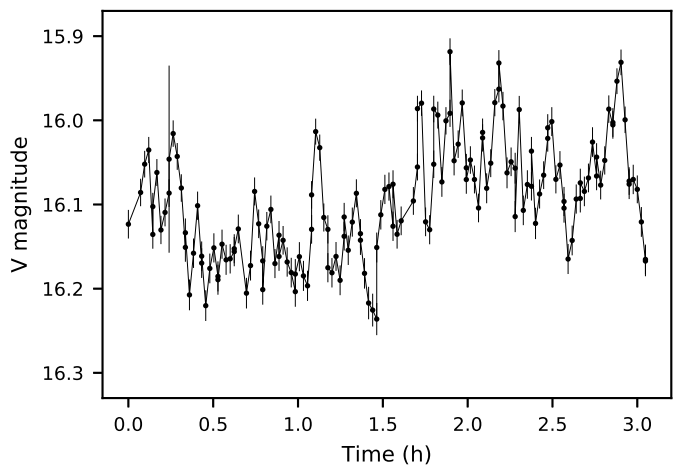}
 \caption{GI Mon TUBITAK light curve, showing variability but no obvious periodicity}
 \label{fig:gimontubi}
\end{figure}

In Figure \ref{fig:gimontubi} we show the TUBITAK light curve of GI~Mon. Although there is obvious short-term variability in the light curve, there is no sign of any periodicity. We confirmed this negative result with an AoV period search. The SED (bottom right panel of Figure \ref{fig:gimon_fullpage}) shows little variability in optical and UV, but a very hard X-ray spectrum consistent with a hot plasma.

\begin{table}
   \caption{Single-temperature X-ray spectral fit for GI~Mon. Fluxes and luminosities are in the 0.2-10.0\,keV range, errors are 1$\sigma$ uncertainties}
   \begin{tabular}{ll}
       nH & $9.7^{+4.4}_{-3.0}\times 10^{20}$\,cm$^{-2}$\\
       kT & $>10$\,keV \\
       norm & $2.9^{+0.4}_{-0.6}\times 10^{-5}$ \\ \hline
       $\chi^2_\nu$ & 1.33 (19)\\ \hline
       Flux   & $ 4.9\pm0.3\times 10^{-14}$\,erg\,cm$^{-2}$\,s$^{-1}$ \\
       Lum    & $ 4.7^{+1.2}_{-1.0}\times 10^{31}$erg\,s$^{-1}$ \\
   \end{tabular}
   \label{tab:gimonspec}
\end{table}

\subsection{HZ Pup}

Our \XMM\ observation of HZ~Pup was 42.8\,ks long. For the EPIC instruments we used source and background extraction regions of 11.1, 22, and 22 arcsec for the PN, MOS1, and MOS2 cameras respectively. The top panel of Figure \ref{fig:hzpup_fullpage} shows the X-ray light curves for all three instruments, background subtracted and binned into intervals of 30\,s in an energy range of 0.2 to 10\,keV.

\begin{figure}
\caption{Periodicities for HZ Pup in X-rays (dashed curve, H-statistic) and UVW1 (solid curve, AoV statistic). Vertical dotted lines indicate the 1210.9\,s spin period and the likely 1292.6\,s $\omega-\Omega$ beat frequency.}
\includegraphics{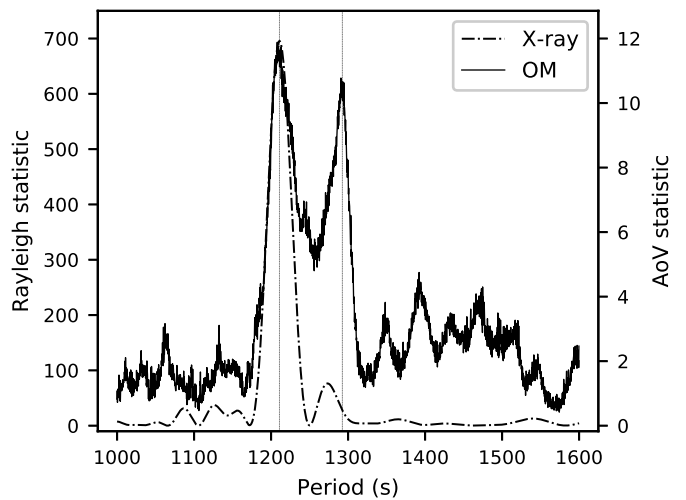}
\label{fig:HZPup_periods}
\end{figure}

The most conspicuous feature of the X-ray light curve of HZ~Pup is the obvious modulation at $\sim$ 20\,min, matching the candidate spin period found by \cite{AbbottShafter1997}. We used the H-test method on the EPIC-{\it pn} data to search for periodicities between 1000 and 1600 seconds to identify the spin period and candidate beat periods identified in that study. We found the spin period to be $1210.90\pm 0.58$\,s, with the uncertainty obtained with the bootstrapping method \citep{Efron1979}. None of the other periods suggested in \cite{AbbottShafter1997} was present in X-rays; see Figure \ref{fig:HZPup_periods}.

We also sought periodicities in the OM light curves, using the AoV method. Two obvious signals are present: the spin period at 1210.9\,s and a slightly weaker signal at 1292.6\,s as shown in Figure \ref{fig:HZPup_periods}. This second periodicity corresponds to the $\omega-\Omega$ (spin frequency minus orbital frequency) beat in \cite{AbbottShafter1997}. The beat frequency is absent in X-rays.

To characterise how the X-ray spectrum of HZ Pup varies with the WD spin we extracted two sets of phase-folded spectra: the bright and faint spin phases, using the interval $0.0<\phi_\text{spin}<0.5$ for the bright phase and $0.58<\phi_\text{spin}<0.92$ for the faint phase; see Figure \ref{fig:hzpup_fullpage}. The spin phase zero point is arbitrary- we used the \XMM\ zero time. 

For the faint phase we found that a single-temperature APEC plasma plus a Gaussian around 6.4\,keV, all subject to a partially covering absorber, gave an excellent fit (see Tab.~\ref{tab:hzpup_xrayspec}). Owing to the faintness of the source, the APEC temperature was unconstrained if left free so we fixed it at 15\,keV.

The same model, allowing for luminosity and covering fraction variation (see sec. \ref{sec:V349Aqr}), does not fit the bright-phase spectrum. Strong residuals below 0.3\,keV, around 1\,keV, and the iron lines indicate that a second plasma component is needed. We held the faint phase spectrum fixed and superimposed a second APEC plus Gaussian model (ie.~{\tt pcfabs$_1$*(apec$_1$+gauss$_1$)+pcfabs$_2$*(apec$_2$+gauss$_2$)}). This new component required a Gaussian line around 0.5\,keV, but no additional iron line emission.

The brightness and width of the iron line Gaussian component in the faint phase suggests that we might be able to resolve the three individual lines. To do this we re-extracted the faint phase spectra with one count per bin and fit them between 5.5 and 7.5\,keV with the C-statistic. The goodness-of-fit was estimated using the method of \cite{Kaastra2017}. We used a bremsstrahlung of fixed temperature 15\,keV to represent the continuum and three Gaussians for the lines. The energies of the He-like and H-like iron lines were fixed to 6.698\,keV and 6.927\,keV, respectively (see Table 1 of \citealt{MeweEtAl1985}) but the energy of the fluorescence line near 6.4\,keV was left variable. The widths and normalisations of all three lines were allowed to vary.

This experiment gave a C-statistic of 402.1 for 496 degrees of freedom. The method described by \cite{Kaastra2017} predicts 403.1 with a standard deviation of 23.7. We have therefore found a statistically good fit. The equivalent width of the fluorescent line was 270\,eV. This value is higher than the 150\,eV typically found in magnetic CVs \citep{Mukai2017}, and higher than most of the IPs studied by \cite{EzukaIshida1999}, but similar to other soft IPs such as NY~Lup \citep{HaberlEtAl2002}, and much lower than the 1.4\,keV found for V902~Mon by \cite{WorpelEtAl2018}. This result suggests that scattering is present but less important in HZ~Pup than for those two sources. Since the iron line emission is constant throughout the WD spin, it is likely that this source accretes at two poles and that the iron line arises from the perpetually visible pole.

\begin{table}
    \caption{Spectral fit for faint and bright spin phases of HZ~Pup. The bright phase consists of an absorbed APEC+gauss component superimposed upon the faint phase spectrum. Fluxes are in the 0.2-10.0\,keV range, and the flux for the bright phase spectrum is only for the additive component. Uncertainties are at the $1\sigma$ level.}
    \begin{tabular}{ll}
     {\bf Faint continuum}\\
        $nH_1$ & $11.6\pm1.1\times 10^{22}$cm$^{-2}$\\
        cvr. fract$_1$ & $95.2\pm0.5\%$ \\
        kT$_1$ & 15\,keV \\
        norm$_1$  & $1.22\pm0.05\times 10^{-3}$\\
        Line energy$_1$ & $6.47\pm0.05$\,keV\\
        Line width$_1$ & $0.19\pm0.04$\,keV \\
        Line norm$_1$ & $1.55_{-0.23}^{+0.22}\times 10^{-5}$\\\hline
        \multicolumn{2}{l}{\bf Additional bright phase emission}\\
        $nH_2$ & $1.45_{-0.28}^{+0.22}\times 10^{22}$cm$^{-2}$\\
        cvr. fract$_2$ & $67.8_{-0.24}^{+0.28}\%$\\
        kT$_2$ & $64_{-7}$\,keV\\
        norm$_2$ & $1.14\pm0.04\times 10^{-3}$ \\
        Line energy$_2$ & $0.521_{-0.015}^{+0.014}$\,keV \\
        Line width$_2$ & $0.106_{-0.014}^{+0.016}$\,keV\\
        Line norm$_2$ & $1.45_{-0.27}^{+0.26}\times 10^{-4}$\\
        $\chi^2_\nu$ (faint phase) & 0.828 (122) \\
        $\chi^2_\nu$ (bright phase) & 1.054 (463) \\\hline
        Flux (faint continuum) & $6.55^{-0.23}_{+0.16}\times 10^{-12}$\,erg\,s$^{-1}$\,cm$^{-2}$\\
        Lum (faint continuum) & $1.6^{+0.4}_{-0.3}\times 10^{33}$\,erg\,s$^{-1}$\\
        Flux (bright component) & $6.16_{-0.09}^{+0.11}\times 10^{-12}$\,erg\,s$^{-1}$\,cm$^{-2}$\\
        Lum (bright component) & $3.9^{+1.1}_{-0.8}\times 10^{33}$\,erg\,s$^{-1}$\\

    \end{tabular}
    \label{tab:hzpup_xrayspec}
\end{table}

The SED for HZ~Pup shows the typical optical/UV behaviour, but its most striking feature is the X-ray portion. After a steady decrease below $2\times 10^{17}$\,Hz, it increases again. This hard X-ray spectrum is again typical of IPs.

\subsection{V597 Pup}

Our \XMM\ observation of V597~Pup was 30.2\,ks long and was unaffected by flaring, except for the last 2\,ks. V597~Pup was very faint in the X-ray data and was only visible after removing the flaring interval, and then only in EPIC-$pn$. Using equation \ref{eq:loredo} with circular source and background regions of 10 and 40 arcsec respectively, we get a count rate of $8.8^{+3.4}_{-3.2}\times 10^{-4}$ counts per second. There were only 60 source region photons, not enough to extract a spectrum, but for a plasma model of temperature 15\,keV this count rate gives a bolometric flux of $7.5^{+2.9}_{-2.7}\times 10^{-15}$\ergx. The distance to the source is not well known (see Table \ref{tab:Gaia}), and its luminosity can be estimated only to the order of $\sim 10^{31}$\,erg\,s$^{-1}$. The source is visible in the OM data, but very faint, at an AB magnitude of $m_\text{UVW1}=21.0$, and we could not get a meaningful light curve or period search from it.

There are eighteen Swift observations of V597~Pup. Of these, all but the last were monitoring it after its 2007 nova eruption. These are not useful for characterising the IP luminosity function because it was still bright, but are potentially useful for finding the spin period. The final Swift observation was taken after the \XMM\ one, but the source is not visible in X-rays or in UV.

To attempt to find the spin period of V597~Pup, we combined the event lists of all the Swift observations except the last into one and performed a H-test on their barycentre-corrected arrival times. There was no signal other than Swift's orbital period and its aliases. Since the source is very faint in the XMM-Newton data, and has little data in optical archives, we did not produce an overview graph for it.

\subsection{V1039 Cen}

The \XMM\ observation was 32\,ks long. It was affected by some periods of slight flaring, but these are not severe enough to adversely affect the data analysis. We used extraction regions of 9.8, 10.6, and 9.2 arcsec for PN, MOS1, and MOS2 respectively.

The source was faint in the X-ray data. We were able to extract a coarse light curve, on 600\,s binning (Figure \ref{fig:v1039cen_fullpage}, top panel). The source appears to be variable, and the phase-folded light curve seems to confirm the 5.9\,hour orbital period suggested by \cite{WoudtEtAl2005} in EPIC$-pn$ (Figure \ref{fig:v1039cen_fullpage}, third panel) but there is no indication of a spin periodicity. V1039~Cen has faded almost to undetectability in UV, with a mean magnitude of $m_\text{UVW1}=21.4$. The light curve hints at some variability but, again, we do not see any periodicity at the proposed spin period. Following the same method as for GI Mon in Section \ref{sec:gimon}, we found that the pulsed fraction is less than 33\% for both the X-rays and UV.

Folding the UV light curve on the orbital period and candidate spin period found in \cite{WoudtEtAl2005} gave no obvious signal due to the source's faintness (Figure \ref{fig:v1039cen_fullpage}, fifth and sixth panels). We extracted an X-ray spectrum of V1039~Cen and obtained a good fit using a partially absorbed APEC model; see Table \ref{tab:v1039censpec} and the bottom left panel of Figure \ref{fig:v1039cen_fullpage}.

\begin{table}
   \caption{Single-temperature X-ray spectral fit for V1039~Cen ({\tt pcfabs*apec}). Fluxes and luminosities are bolometric, errors are 1$\sigma$ uncertainties}
   \begin{tabular}{ll}
       nH & $1.1^{+0.4}_{-0.3}\times 10^{22}$\,cm$^{-2}$\\
       cvrFrac & 0.94$\pm 0.04$ \\
       kT & $7.9_{-2.6}^{+5.1}$\,keV \\
       norm & $6.7^{+1.2}_{-0.8}\times 10^{-5}$ \\ \hline
       $\chi^2_\nu$ & 0.87 (40)\\ \hline
       Flux   & $ 9.6\pm0.6\times 10^{-14}$\,erg\,cm$^{-2}$\,s$^{-1}$ \\
       Luminosity    & $ 5.6^{+8.1}_{-3.2}\times 10^{31}$erg\,s$^{-1}$ \\
   \end{tabular}
   \label{tab:v1039censpec}
\end{table}

The SED for this source is very sparse owing to its faintness and the highly uncertain distance. We did not observe any obvious stellar component attributable to the donor, and the X-ray energy distribution rises steadily with increasing frequency, indicating that this is a hard X-ray source.

\subsection{V1084 Her}

The \XMM\ observation of V1084~Her was 24\,ks long. The observation was unaffected by soft proton flaring. We extracted source region photons from circles of radius 15.9, 18.4, and 22 arcsec from PN, MOS1, and MOS2 respectively.

The X-ray light curve (see top panel of Figure \ref{fig:v1084her_fullpage}) shows slight orbital modulation at the 2.9\,hr period found by \cite{MickaelianEtAl2002} as well as intense shorter-term variability. The orbital modulation is more clear in the phase folded panel. In the OM light curve the orbit is also clearly visible, and there is also an indication of shorter term variability. Our period searches clearly revealed the orbital period, and also found slight evidence for a shorter signal at 1,549\,s in both the X-rays and in the UV. The peaks and troughs of the two light curves do not coincide, however, and the light curves folded on this period are only convincing in the UV, so this 26\,min signal can only be regarded as tentative. Furthermore, the signal was only visible in the higher energy half of the X-ray photon list, so associating it with the spin period is dubious. The overall phase-folded light curve was distinguishable from a constant with a probability of about $10^{-2}$, which is only a marginal detection.

We were able to get a decent fit to the \XMM\ X-ray data using a spectrum consisting of a partially absorbed two-temperature APEC model plus a Gaussian near 6.4\,keV; see Figure \ref{fig:v1084her_fullpage}. The width of the Gaussian was fixed at effectively zero. Our best-fit spectral parameters are given in Table \ref{tab:v1084herspec}.

\begin{figure}
    \includegraphics{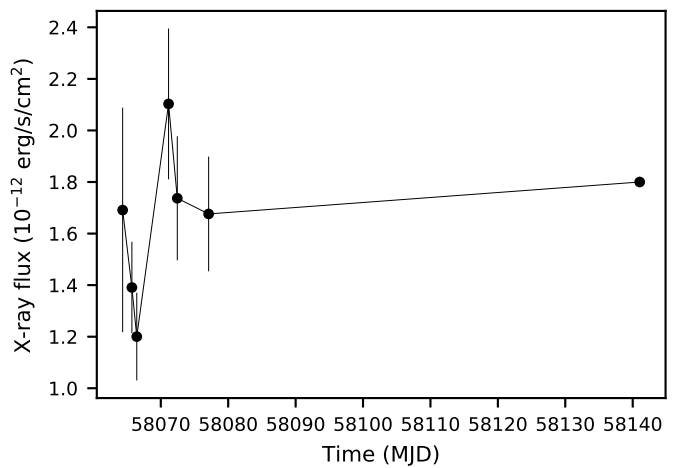}
    \caption{Long-term X-ray light curve of V1084 Her. The first six observations were performed by Swift and the last by XMM-Newton}
    \label{fig:V1084Her_xraylicu}
\end{figure}

\begin{table}
\caption{Spectral fit for V1084 Her ({\tt pcfabs*(apec + apec + gaussian}). Fluxes and luminosities are bolometric, errors are 1$\sigma$ uncertainties}
\begin{tabular}{ll}
nH & $1.2^{+0.2}_{-0.1}\times 10^{22}$\,cm$^{-2}$ \\
cvr. fract & $0.61\pm 0.03$ \\
kT$_1$ & $0.93^\pm 0.03$\,keV  \\
norm$_1$ & $9.7\pm 1.3\times 10^{-5}$  \\ 
kT$_2$ & $6.1\pm0.4$\,keV \\
norm$_2$ & $8.3\pm0.3\times 10^{-4}$ \\
Line energ. & $6.36^{+0.02}_{-0.03}$\,keV \\
Line norm   & $2.8\pm0.7\times 10^{-6}$ \\ \hline
$\chi^2_\nu$ & 1.22 (498) \\ \hline
abs. flux   & $ 1.80\pm0.02\times 10^{-12}$\,erg\,cm$^{-2}$\,s$^{-1}$ \\
abs. lum    & $ 4.05\pm0.09\times 10^{31}$erg\,s$^{-1}$ \\
\end{tabular}
\label{tab:v1084herspec}
\end{table}
The source was visible in all six Swift observations, so we used the method described in Sect. \ref{sec:swift} to track its X-ray luminosity. We found that the X-ray flux of the source is mildly variable, ranging from $1.2-2.1\times 10^{-12}$\,erg\,s$^{-1}$\,cm$^{-2}$; see Figure \ref{fig:V1084Her_xraylicu}.

The spectral energy distribution (see bottom right panel of Fig. \ref{fig:v1084her_fullpage}) shows, as usual, the stellar behaviour at long wavelengths. The variability in optical and UV is very pronounced, and the X-ray region becomes flat above $\nu>10^{17}$\,Hz and may rise again at $10^{18}$\,Hz.

\subsection{IGR~J18151-1052}

This object was very bright in X-rays but almost undetectable in UV. The X-ray light curve (top panel of Figure \ref{fig:igr_fullpage}) shows obvious and intense variability on timescales of minutes, whereas the UV light curve contains very little data.

In the X-ray data, we found a very conspicuous period of 390.5\,s. Given the length of the observation, 39.9\,ks, we would have been sensitive to an orbital period of up to two or three hours, but we found no significant longer period. Although IPs are frequently found to have an orbital period of the order of ten times longer than the spin, it is not always visible in X-rays (for example \citealt{ParkerEtAl2005}).

As for DQ~Her we then divided the photon lists into high and low energy halves and performed the H-test again on both lists. The division was at 4.505\,keV. This is at a much higher energy than the other sources for which we performed this test because of the significantly higher absorption. Even so, we recovered the 390\,s signal in both energy ranges.

We extracted an X-ray spectrum for this source and fit it with a two-temperature APEC plasma, with extra emission at the 6.4\,keV iron line, modified by one absorber for interstellar material and a partially covering absorber representing local material ({\tt phabs*pcfabs*(apec + apec + gaussian}); see Figure \ref{fig:igr_fullpage} and Table \ref{tab:IGRJ_specfit}. We found that we could not omit either of the absorbers, or either of the plasma components, and still obtain an adequate fit. Owing to its proximity to the Galactic plane, and therefore significant absorption, we do not have enough optical data points to make an SED worthwhile.

\begin{table}
   \caption{Spectral fit for IGR~J18151-1052 ({\tt phabs*pcfabs*(apec + apec + gaussian}). Errors are at the $1\sigma$ level and fluxes are in the 0.2-10.0\,keV range.}
   \begin{tabular}{lc}
       nH$_1$ & 1.25$^{+0.04}_{-0.03}\times 10^{22}$\,cm$^{-2}$\\
       nH$_2$ & 15.7$\pm0.9\times 10^{22}$\,cm$^{-2}$\\
       Cover Fract & 76.5$^{+0.7}_{-1.1}$\% \\
       kT$_1$ & $>46.6 $\,keV\\
       norm$_1$ & 2.16$^{+0.27}_{-0.19}$ \\
       kT$_2$ & $8.66_{-0.72}^{+0.77}$\,keV\\
       norm$_2$ & 1.20$^{+0.10}_{-0.11}$ \\
       Line energ. & $6.41\pm0.01$\,keV \\
       Line norm & $1.08\pm 0.11 \times 10^{-5}$\\
       Line eq. wd. & $223\pm 23$\,eV\\ \hline
       $\chi^2_\nu$ & 0.92 (777) \\
       Absorbed flux & 1.18$\pm 0.01\times 10^{-13}$\ergx \\
   \end{tabular}
   \label{tab:IGRJ_specfit}
\end{table}

\subsection{V4745 Sgr}

The \emph{Swift} observation shows V4745~Sgr visible in 2007 Aug, four years after its nova eruption, but it had faded to undetectability in X-rays by 2018 for our 38.2\,ks observation. To estimate its flux we assumed a 15\,keV APEC spectrum with interstellar absorption of $9.0\times 10^{20}$\,cm$^{-2}$ \citep{SchwarzEtAl2011} and used the {\tt fakeit} command of Xspec to produce a simulated spectrum. Then simple photon counting gives a flux of $3.6\pm0.7\times 10^{-13}$\,\ergx\ for the Swift observation and an upper limit of $2\times 10^{-15}$\,\ergx\ for the \XMM\ observation assuming we would have required ten source photons to identify it by eye. At its furthest plausible distance of 8.4\,kpc (see Table \ref{tab:Gaia}) the maximum X-ray luminosity is $2\times 10^{31}$\,\ergs.

\begin{figure}[h!]
    \includegraphics{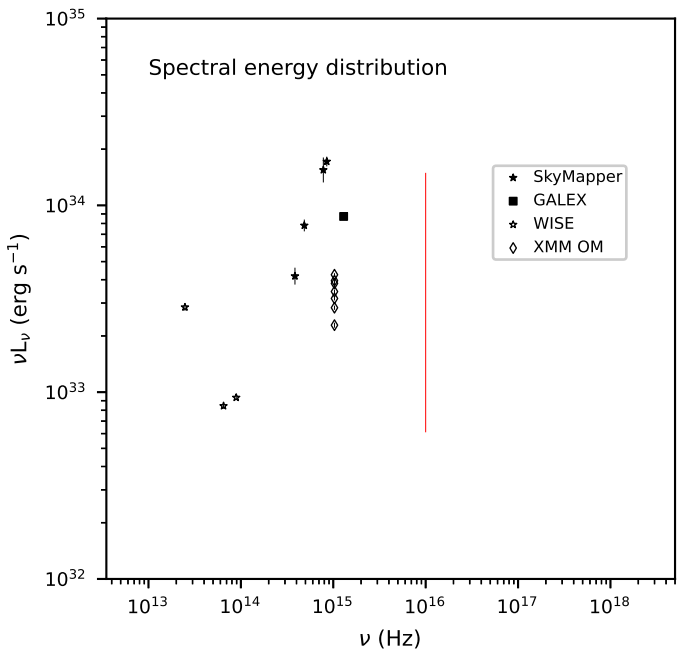}
    \caption{Spectral energy distribution for V4745 Sgr. The vertical red line indicates the uncertainty in luminosity arising from the uncertainty in the distance.}
    \label{fig:V4745Sgr_sed}
\end{figure}

It was still visible in UV in the \XMM\ observation as a faint source of $m_{UVW1}=18.9$. We have only enough data to produce a partial SED, which is shown in Figure \ref{fig:V4745Sgr_sed}.

\subsection{V533 Her}

We have one XMM-Newton observation of this source, lasting 29.4\,ks, and it is not significantly affected by periods of high background. The H-test applied to the arrival times of photons from all X-ray instruments showed no obviously significant periodicity; there is only a marginal indication of signal around 1349\,s (see Figure \ref{fig:V533Her_fullpage}). Variability is seemingly detected in EPIC-$pn$ at that period; the null hypothesis has probability less than one part in 3,000.

There was no sign of the 1,400\,s period found by \cite{Rodriguez-GilMartinez-Pais2002} or the former 63.6\,s periodicity \citep{Patterson1979}. A pulsed fraction of 10\% would have been detectable in X-rays.

The OM light curve (see Figure \ref{fig:V533Her_fullpage}) clearly shows the 3.53\,hr orbital modulation of the system but there is no sign of this feature in X-rays. Nor is there any evidence for the previously suggested spin periods at 63.6 and 1400\,s. The possible 1,349\,s feature is not present in optical even after subtracting the orbital modulation from it.

We extracted an X-ray spectrum for the entire observation, for all three instruments, and fit the spectra with an absorbed two-temperature APEC  ({\tt phabs*(apec+apec)}). The results are shown in Table \ref{tab:V533Her_specfit}. The SED for this source shows it to be very luminous optically, but its X-ray luminosity is relatively low and it lacks any hint of a soft component.

\begin{table}
   \caption{Spectral fit for V533 Her ({\tt phabs*(apec+apec)}). Uncertainties are at the $1\sigma$ level and flux is in the 0.2-10.0\,keV range.}
   \begin{tabular}{lc}
       nH & 7.2$^{+0.09}_{-0.21}\times 10^{21}$\,cm$^{-2}$\\
       kT$_1$ & $0.156^{+0.026}_{0.023}$\,keV\\
       norm$_1$ & 1.8$^{+4.9}_{-1.5}\times 10^{-3}$ \\
       kT$_2$ & $9.4^{+4.8}_{-1.7}$\,keV\\
       norm$_2$ & 3.2$\pm 0.1\times 10^{-4}$ \\
       $\chi^2_\nu$ & 0.944 (224) \\
       Absorbed flux & 5.1$\pm 0.1\times 10^{-13}$\ergx \\
       Absorbed luminosity & 8.3$\pm 0.5\times 10^{31}$\ergs \\
   \end{tabular}
   \label{tab:V533Her_specfit}
\end{table}

\subsection{V1425 Aql}

This source V1425~Aql was very faint in the XMM-Newton observation, and we could extract only a very coarse spectrum from the EPIC-$pn$ and EPIC-MOS1 instruments. The source was not visible in MOS2. The spectrum was adequately fitted with an absorbed one-temperature APEC plasma; see Table \ref{tab:V1425Aql_specfit}. We did not have enough photons to perform a meaningful period search or to construct light curves.

\begin{table}
   \caption{Spectral fit for V1425~Aql ({\tt phabs*apec}). Fluxes are in the 0.2 to 10.0 \,keV range and  uncertainties are $1\sigma$.}
   \begin{tabular}{lc}
       nH & 11.7$^{+3.1}_{-2.4}\times 10^{22}$\,cm$^{-2}$\\
       kT & $6.9^{+4.0}_{2.2}$\,keV\\
       norm$_2$ & 8.2$^{+2.5}_{-1.7}\times 10^{-5}$ \\
       $\chi^2_\nu$ & 0.84 (20) \\
       Flux & 5.7$\pm 0.6\times 10^{-14}$\ergx \\
   \end{tabular}
   \label{tab:V1425Aql_specfit}
\end{table}

The optical counterpart to the 1995 nova V1425~Aql is a faint bluish object at 19:05:26.73 -1:42:4.2 \citep{NakanoEtAl1995}, approximately 1.7" northwest of a brighter star found in Palomar plates by \cite{Skiff1995}. It is not in the Gaia DR2 catalogue, so we do not have a distance estimate for it, and there are only $r$ and $z$ magnitudes visible in PanSTARRS (PSO~J190526.729-014204.142).

There was no detectable periodicity in the OM data. V1425~Aql was not visible in any of the six \emph{Swift} observations, with either the XRT or the UVOT. Owing to the lack of data in other wavebands, we have omitted the full-page overview of this source in the Appendix.

\subsection{V349 Aqr}
\label{sec:V349Aqr}

We have one XMM-Newton observation, 34.8\,ks long, and three Swift observations of this source. There were two periods of moderately elevated background but these were not severe enough to significantly affect the spectra. The Swift pointings total 4.1\,ks and were acquired a few years before the XMM observation.

The H-test period search applied to the barycentre-corrected EPIC-pn source region event list gives a very strong signal at 390.15\,s. This period is very likely the IP spin period, even though \cite{WoudtEtAl2004} proposed this to be the $1/{P_{spin}}+1/{P_{orb}}$ beat period. The stronger optical period at 403.7 s (proposed as the spin period by \citealt{WoudtEtAl2004}), then, is in fact the $1/{P_{spin}}+1/{P_{orb}}$ beat period. The orbital period, inferred from the separation of these periods, $P_{orb}=3.23$\,hr, remains unchanged. The spin-folded phase-folded light curve is shown in Figure \ref{fig:v349aqr_fullpage}. The AoV method applied to the OM data gave a hint of the 390\,s periodicity, but the phase-folded light curve is consistent with no variability with $70\%$ probability. 

As for DQ~Her we divided the $pn$ photon list into high and low energy halves- for this source the division was at 1.8\,keV. The 390\,s signal was much stronger for the low energy half than for the high energy half, although it was significantly detected in both. The orbital period of 3.23 hours inferred by \cite{WoudtEtAl2004} was not detected in any of our period searches, at either X-ray or UV wavelengths. Neither was the tentative 2.0 hour period suggested from spectroscopy by \cite{SzkodyEtAl2003}.

To characterise the X-ray spectrum of V349~Aqr, we extracted bright and faint spin phase spectra. We then fitted the faint spectra jointly with a partially absorbed two-temperature plasma ({\tt pcfabs*(apec+apec)}). We used the same spectral parameters for the bright and faint spectra, except that we allowed the normalisations of the plasma components and the plasma covering fraction to vary between them. This differs from the approach taken for HZ~Pup, which did not give a good fit in this case. The results are given in Table \ref{tab:V349_specfit}. V349~Aqr has an X-ray luminosity relatively bright compared to its optical flux (see SED in Figure \ref{fig:v349aqr_fullpage}) and requires a two temperature APEC models to achieve a good fit.

\begin{table*}
   \caption{Spectral fit for V349~Aqr ({\tt pcfabs*(apec+apec)}). Fluxes are in the 0.2 to 10.0 \,keV range and  uncertainties are $1\sigma$.}
\begin{tabular}{lcc}
                       & faint                                & bright \\
   nH                  & $1.31^{+0.34}_{-0.29}\times 10^{22}$\,cm$^{-2}$ &\\
   Covering fraction   & $73.2^{+3.8}_{-3.5}$\% & $53.6^{+5.6}_{-3.6}$\%\\
   kT$_1$              & $1.32^{+0.10}_{-0.08}$\,keV &\\
   norm$_1$            & $2.91^{+1.50}_{-1.14}\times 10^{-5}$ & $3.08^{+1.18}_{-0.95}\times 10^{-5}$\\
   kT$_2$              & $>32$\,keV &\\
   norm$_2$            & $3.94^{+0.18}_{-0.42} \times 10^{-4}$ & $6.37^{+0.21}_{-0.63} \times 10^{-4}$\\ \hline
   $\chi^2_\nu$        & 1.01 (339) &\\ \hline
   Abs. flux           & $7.8\pm 0.2\times 10^{-13}$\,erg\,s$^{-1}$\,cm$^{-2}$ & $13.3\pm 0.2\times 10^{-13}$\,erg\,s$^{-1}$\,cm$^{-2}$\\
   Abs. Lum.           & $4.5^{+2.6}_{-1.7}\times 10^{32}$\,erg\,s$^{-1}$& $7.7^{+4.5}_{-3.0}\times 10^{32}$\,erg\,s$^{-1}$\\
\end{tabular}
\label{tab:V349_specfit}
\end{table*}

\begin{figure}
    \includegraphics{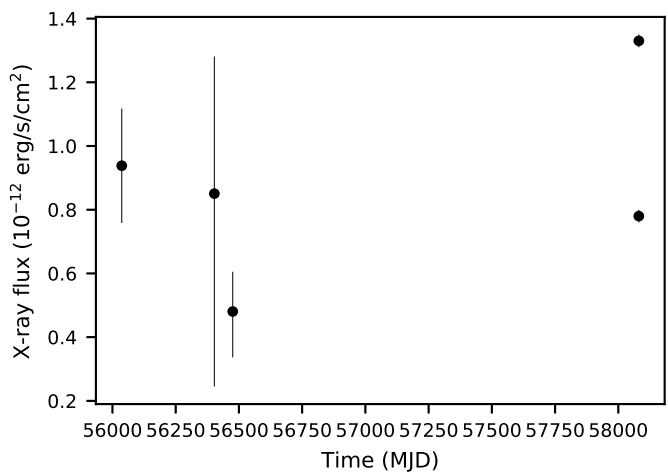}
    \caption{Long-term X-ray light curve of V349~Aqr. The first three points are Swift observations; the final two points are the faint and bright phase XMM points as described in Table \ref{tab:V349_specfit}.}
    \label{fig:V349Aqr_xraylicu}
\end{figure}

The source is visible in all three \emph{Swift} observations, suggesting that it is a persistently active source. None of the observations were long enough to give enough photons for a spectral fit, so we estimated the source's flux as described in Sect \ref{sec:swift}. The results are shown in Figure \ref{fig:V349Aqr_xraylicu}. The source is clearly variable.

\section{Conclusion}

We confirmed that HZ~Pup is an IP. Its spin modulation of 1552\,s in X-rays is indisputable. Its phase-averaged bolometric luminosity is around $6.8\times 10^{33}$\,erg\,s$^{-1}$ so it is a typical IP and not one of the sought-after X-ray underluminous IPs.

Our results confirm that V349~Aqr is an IP. We have found a 390\,s periodicity in X-rays that demonstrates that the 404\,s signal earlier found optically is the beat period. We have therefore confirmed distinct spin and sideband periods in optical and X-rays. The X-ray luminosity of this object is $5.7\times 10^{32}$\,\ergs. This value makes it one of the typical X-ray bright IPs, and not one of the X-ray underluminous population.

The highly absorbed object IGR J18151-1052 has a significant periodicity at 390.8\,s. This may be evidence of WD spin, but the source is so faint optically that identifying its nature is challenging. If it is an intermediate polar, as seems likely, its X-ray flux is comparable to those of the normal members of the class at a distance of several kiloparsecs. Owing to its optical faintness, we do not have a Gaia distance for it nor do we know anything about the companion star, so we cannot currently characterise its X-ray luminosity. Further observations aimed at discovering its orbital period would be helpful.

Although several of our targets have periodicities of around 390\,s, this is unlikely to be a systematic effect. We tested this by performing period searches on areas of background in these observations, and other sources in the field, and in no case did we find the 390\,s signal. Additionally, this period is consistent with the beat period found for V349 Aqr. The signal found in our sources is therefore real and their similarity is coincidental. Interestingly, the previously known intermediate polar CC~Scl also has a spin period of 390\,s \citep{WoudtEtAl2012}.

Of the remaining targets, GI~Mon showed plasma emission but neither spin nor orbital periods were evident. It is therefore probably not a magnetic CV, although its strong short-term optical variability makes it an interesting source for follow-up observations. The southern source V1039~Cen was faint owing to its great distance, but the orbital modulation was clear in X-rays. The lack of a shorter periodicity attributable to a distinct spin period, as well as the somewhat low plasma temperature, suggests that it is a polar magnetic CV rather than an IP but we cannot rule out the latter possibility.

V1084 Her is an interesting object. The orbital period is clear in both X-rays and UV, and there are inconclusive hints of a shorter periodicity of around 26 minutes. This signal is not obviously attributable to a rapid WD spin, and the plasma temperature is low enough to suggest that, if magnetic at all, this object is also a polar CV rather than an IP. However, the blue continuum and emission line properties \citep{MickaelianEtAl2002} and the presence of what seems to be a negative superhump in the photometry \citep{PattersonEtAl2002} suggest the presence of an accretion disc. The circular polarisation should also be modulated on the spin period, which was found not to be the case by \cite{RodriguezGilEtAl2009}. We therefore conclude the object is probably not a magnetic CV.

The former nova V1425 Aql has now faded almost to undetectability and we were unable to confirm either an orbital or a spin period. This source may not be a magnetic CV at all.

The former nova V533~Her showed clear orbital modulation at optical/UV wavelengths but not in X-rays. Similarly, we found a possible signal in X-rays at 1349\,s, but this could not be detected in the OM data. The classification of V533~Her as an IP is therefore still plausible, but could not be confirmed here. Its SED indicates a hard but faint X-ray spectrum, suggestive of a low X-ray luminosity IP. V533~Her remains a promising candidate for future observations.

Among our sample, we were able to confirm the IP nature for two systems with high amplitude short period optical oscillations well above 2\%: HZ Pup ($>40 $\,mmag; \citealt{AbbottShafter1997}) and V349~Aqr ($>30$\,mmag; \citealt{WoudtEtAl2004}). For the highly obscured IGR~J18151-1052 we have no good optical photometry. These sources showed X-ray spin period modulations of the order of 50\%. Empirically, therefore, such high amplitude optical modulations are good indicators of their IP nature. For lower amplitude modulations, which may represent flickering or QPOs, we advocate a more cautious approach, to verify the persistence and coherence of such periods before accepting the sources as likely IPs.

Of the remaining unconfirmed candidates, only V597 Pup and V4745 Sgr have luminosities at or below the $10^{31}$\,\ergs\ necessary to consider them possible members of the underluminous IP class (the latter may be very much fainter). Further observations of these objects would be helpful.

\begin{acknowledgements}
This work was supported by the German DLR under contracts 50 OR 1405, 50 OR 1711, and 50 OR 1814. Samet Ok is supported by TUBITAK 2214-A International Doctoral Research Fellowship Programme. We thank TUBITAK for a partial support in using the T100 telescope with project number 16BT100-1027. We acknowledge with thanks the variable and comparison star observations from the AAVSO International Database contributed by observers worldwide and used in this research. Many of the AAVSO observations were the result of Centre for Backyard Astronomy (CBA) campaigns. This research has made use of the APASS database, located at the AAVSO web site. Funding for APASS has been provided by the Robert Martin Ayers Sciences Fund. This work has made use of Astropy \citep{Astropy2013,Astropy2018}.
This research has made use of data, software and/or web tools obtained from the High Energy
Astrophysics Science Archive Research Center (HEASARC), a service of the Astrophysics Science
Division at NASA/GSFC and of the Smithsonian Astrophysical Observatory's High Energy Astrophysics Division.
 
The national facility capability for SkyMapper has been funded through ARC LIEF grant LE130100104 from the
Australian Research Council, awarded to the University of Sydney, the Australian National University, Swinburne University of Technology, the University of Queensland, the University of Western Australia, the University of Melbourne, Curtin University of Technology, Monash University and the Australian Astronomical Observatory. SkyMapper is owned and operated by The Australian National University's Research School of Astronomy and Astrophysics. The survey data were processed and provided by the SkyMapper Team at ANU. The SkyMapper node of the All-Sky Virtual Observatory (ASVO) is hosted at the National Computational Infrastructure (NCI). Development and support the SkyMapper node of the ASVO has been funded in part by Astronomy Australia Limited (AAL) and the Australian Government through the Commonwealth's Education Investment Fund (EIF) and National Collaborative Research Infrastructure Strategy (NCRIS), particularly the National eResearch Collaboration Tools and Resources (NeCTAR) and the Australian National Data Service Projects (ANDS).

We are grateful to the anonymous referee, whose comments led to large improvements in the clarity of the paper.

\end{acknowledgements}

\appendix

\section{X-ray Observation log}

Sources are ordered by increasing right ascension, and in chronological order for each source.
The lengths of the observations are those of the X-ray instruments; the exposures of the BAT instrument
on \Swift\ and the UV telescopes on both satellites may be slightly different.

\begin{table}[h!]
\caption{Observation log. Sources are given in order of right ascension, and observations
are ordered chronologically per source.} 
\begin{tabular}{llllr}
 \textbf{Source} & \textbf{Obsid} & \textbf{Date} & \textbf{inst} & \textbf{length}\\
                 &               &                &               & \textbf{  (ks) }\\
 \hline
 GI Mon           & 0804110401  & 2018-04-16 & XMM   & 35.0 \\
 \hline
 HZ Pup           & 0804110501  & 2018-04-21 & XMM   & 42.8 \\
 \hline
 V597 Pup         & 00031024001 & 2007-11-20 & Swift &  1.8 \\
 V597 Pup         & 00031024002 & 2007-12-11 & Swift &  1.9 \\
 V597 Pup         & 00031024003 & 2008-01-08 & Swift &  1.9 \\
 V597 Pup         & 00031024004 & 2008-01-17 & Swift &  4.1 \\
 V597 Pup         & 00031024005 & 2008-01-24 & Swift &  4.9 \\
 V597 Pup         & 00031024006 & 2008-02-17 & Swift &  2.5 \\
 V597 Pup         & 00031024007 & 2008-03-13 & Swift &  1.8 \\
 V597 Pup         & 00031024008 & 2008-04-27 & Swift &  2.2 \\
 V597 Pup         & 00031024009 & 2009-01-06 & Swift &  3.7 \\
 V597 Pup         & 00031024010 & 2009-01-14 & Swift &  6.6 \\
 V597 Pup         & 00031024011 & 2009-01-15 & Swift &  6.7 \\
 V597 Pup         & 00031024012 & 2009-01-27 & Swift &  1.0 \\
 V597 Pup         & 00031024013 & 2009-02-10 & Swift &  1.7 \\
 V597 Pup         & 00031024014 & 2009-02-24 & Swift &  1.3 \\
 V597 Pup         & 00031024015 & 2009-03-11 & Swift &  1.5 \\
 V597 Pup         & 00031024016 & 2009-03-24 & Swift &  1.2 \\
 V597 Pup         & 00090252001 & 2009-06-20 & Swift &  5.1 \\
 V597 Pup         & 0804110901  & 2018-04-07 & XMM   & 30.2 \\
 V597 Pup         & 03100352001 & 2018-10-13 & Swift &  1.1 \\
 \hline
 V1039 Cen        & 0804110201  & 2018-03-07 & XMM   & 32.0 \\
 \hline
 V1084 Her        & 00087575001 & 2017-11-07 & Swift &  0.7 \\
 V1084 Her        & 00087575002 & 2017-11-08 & Swift &  3.2 \\
 V1084 Her        & 00087575003 & 2017-11-09 & Swift &  3.4 \\
 V1084 Her        & 00087575004 & 2017-11-14 & Swift &  1.7 \\
 V1084 Her        & 00087575005 & 2017-11-15 & Swift &  2.2 \\
 V1084 Her        & 00087575007 & 2017-11-19 & Swift &  2.5 \\
 V1084 Her        & 0804110101  & 2018-01-22 & XMM   & 24.0 \\
 \hline
 DQ Her           & 0804111201  & 2017-04-19 & XMM   & 41.9 \\
 \hline
 V533 Her         & 0804110301  & 2017-05-17 & XMM   & 29.4 \\
 \hline
 IGR J18151       & 0823310801  & 2019-03-18 & XMM   & 39.9 \\
 \hline
 V4745 Sgr        & 00035220001 & 2007-08-05 & Swift &  4.8 \\
 V4745 Sgr        & 0804110701  & 2018-04-06 & XMM   & 38.2 \\
 \hline
 V1425 Aql        & 00045873001 & 2012-11-15 & Swift &  0.3 \\
 V1425 Aql        & 00045873002 & 2014-12-01 & Swift &  0.6 \\
 V1425 Aql        & 00045873003 & 2015-02-21 & Swift &  0.3 \\
 V1425 Aql        & 00045873004 & 2015-11-23 & Swift &  0.2 \\
 V1425 Aql        & 00045873005 & 2016-03-16 & Swift &  0.3 \\
 V1425 Aql        & 0804110601  & 2017-10-23 & XMM   & 40.3 \\
 V1425 Aql        & 00045873006 & 2018-12-02 & Swift &  2.6 \\
 \hline
 V349 Aqr         & 00045759001 & 2012-04-19 & Swift &  1.9 \\
 V349 Aqr         & 00045759002 & 2013-04-20 & Swift &  0.3 \\
 V349 Aqr         & 00045759003 & 2013-07-02 & Swift &  1.9 \\
 V349 Aqr         & 0804110801  & 2017-11-24 & XMM   & 34.8 \\
 \end{tabular}
 \label{tab:obslog}
\end{table}

\section{Optical telescope filter effective wavelengths and zero points}

\begin{table}
 \caption{Optical telescope filter effective wavelengths and zero points}

\begin{tabular}{llrrc}
Filter & Inst & Eff $\lambda$ (\AA) & Zero   & refs \\
       &      &                     & point (Jy)  & \\
\hline

FUV      & GALEX      &   1539    &  515     & 1       \\
W2       & Swift-UVOT &   1928    & 3631     & 2       \\
W2       & XMM-OM     &   2210    & 3631     & 3       \\
M2       & Swift-UVOT &   2246    & 3631     & 2       \\
M2       & XMM-OM     &   2310    & 3631     & 3       \\
NUV      & GALEX      &   2316    &  782     & 1       \\
W1       & Swift-UVOT &   2600    & 3631     & 2       \\
W1       & XMM-OM     &   2910    & 3631     & 3       \\
U        & XMM-OM     &   3440    & 3631     & 3       \\
U        & Swift-UVOT &   3465    & 3631     & 2       \\
u        & SkyMapper  &   3525    & 3631     & 4       \\
U        & SDSS       &   3557    & 3631     & 5       \\
v        & SkyMapper  &   3840    & 3631     & 4       \\
B        & XMM-OM     &   4340    & 3631     & 3       \\
B        & AAVSO      &   4361    & 3631     & N/A     \\
B        & Swift-UVOT &   4392    & 3631     & 2       \\
G        & PanSTARRS  &   4810    & 3631     & 6       \\
G        & SDSS       &   4825    & 3631     & 5       \\
g        & SkyMapper  &   5100    & 3631     & 4       \\
White    & CRTS       &   5400    & 3631     & N/A     \\
V        & XMM-OM     &   5430    & 3631     & 3       \\
V        & AAVSO      &   5448    & 3631     & N/A     \\
V        & Swift      &   5468    & 3631     & 2       \\
r        & SkyMapper  &   6170    & 3631     & 4       \\
R        & PanSTARRS  &   6170    & 3631     & 6       \\
R        & SDSS       &   6254    & 3631     & 5       \\
R        & AAVSO      &   6407    & 3631     & N/A     \\
I        & PanSTARRS  &   7520    & 3631     & 6       \\
I        & SDSS       &   7668    & 3631     & 5       \\
i        & SkyMapper  &   7790    & 3631     & 4       \\
I        & AAVSO      &   7980    & 3631     & N/A     \\
Z        & PanSTARRS  &   8660    & 3631     & 6       \\
Z        & VISTA      &   8770    & 3631     & 7       \\
Z        & SDSS       &   9114    & 3631     & 5       \\
z        & SkyMapper  &   9160    & 3631     & 4       \\
Y        & PanSTARRS  &   9620    & 3631     & 6       \\
Y        & VISTA      &  10200    & 3631     & 7       \\
J        & 2MASS      &  12350    & 1594     & 8       \\
J        & VISTA      &  12520    & 3631     & 7       \\
H        & VISTA      &  16450    & 3631     & 7       \\ 
H        & 2MASS      &  16620    & 1024     & 8       \\
K$_S$    & VISTA      &  21470    & 3631     & 7       \\
K        & 2MASS      &  21590    & 666.8    & 8       \\
W1       & WISE       &  33680    & 306.682  & 9       \\
W2       & WISE       &  46180    & 170.663  & 9       \\
W3       & WISE       & 120820    & 29.045   & 9       \\
\end{tabular}

 \tablebib{  1. \cite{MorrisseyEtAl2007, MartinEtAl2005}, 2. \cite{PageEtAl2014, PooleEtAl2008}, 
             3. \cite{KirschEtAl2004}, 4. \cite{WolfEtAl2018},
             4. \cite{FukugitaEtAl1996}, 5. \cite{TonryEtAl2012}, 
             6. \url{http://casu.ast.cam.ac.uk/surveys-projects/vista/technical/filter-set},
             7. \cite{CohenEtAl2003}, 8. \cite{JarrettEtAl2011}}
\label{tab:filters}
\end{table}

\section{Figures}

\begin{figure*}[h!]
    \includegraphics{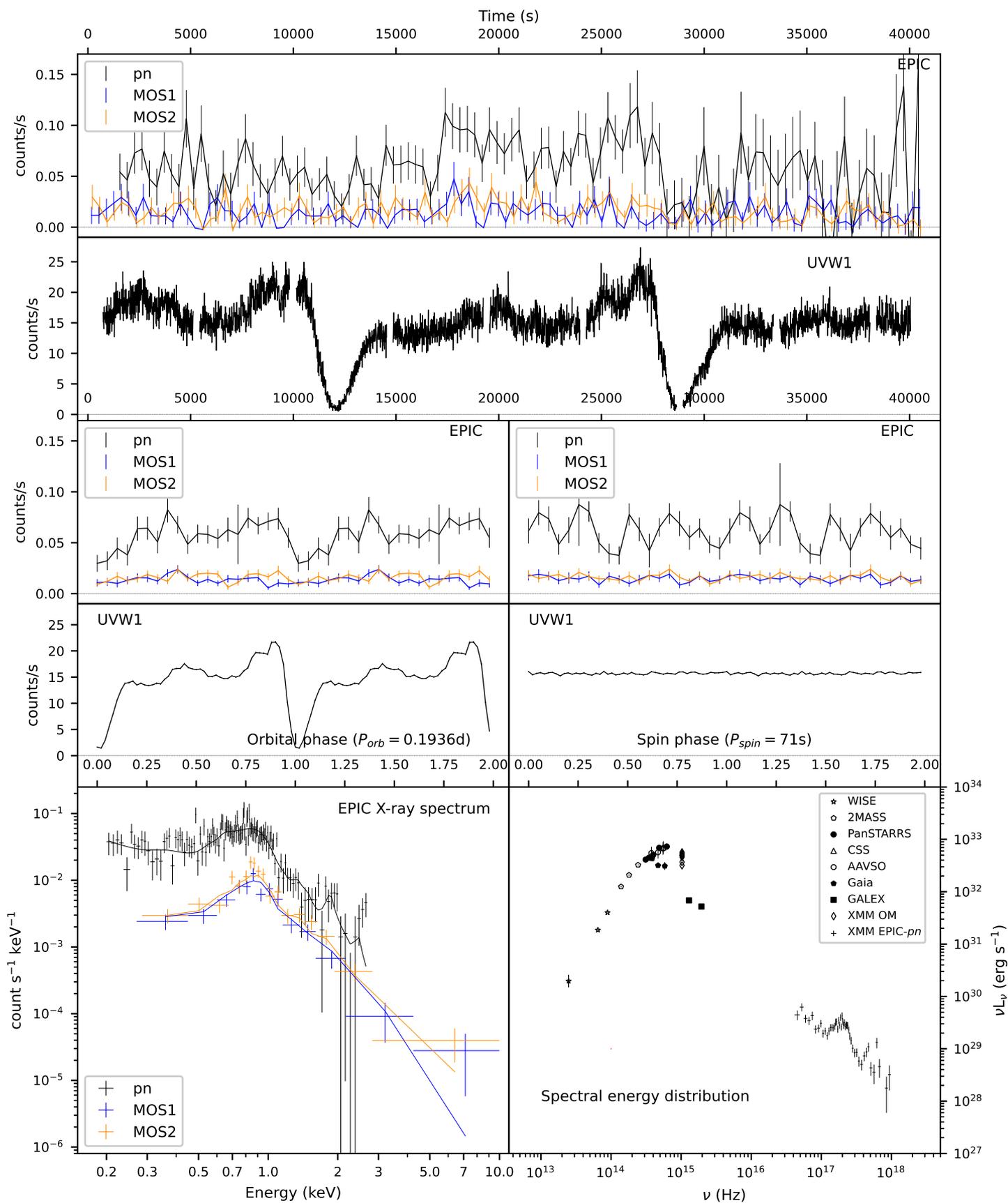}
    \caption{{\it Top panels: }X-ray and UV light curves of DQ~Her; {\it Middle panels:} phase folded on orbital and spin periods; {\it Bottom panels:} X-ray spectrum and SED. The vertical red line here, and in the subsequent Appendix figures, indicates the uncertainty in luminosity arising from the uncertainty in the distance.}
    \label{fig:dqher_fullpage}
\end{figure*}

\begin{figure*}[h!]
    \includegraphics{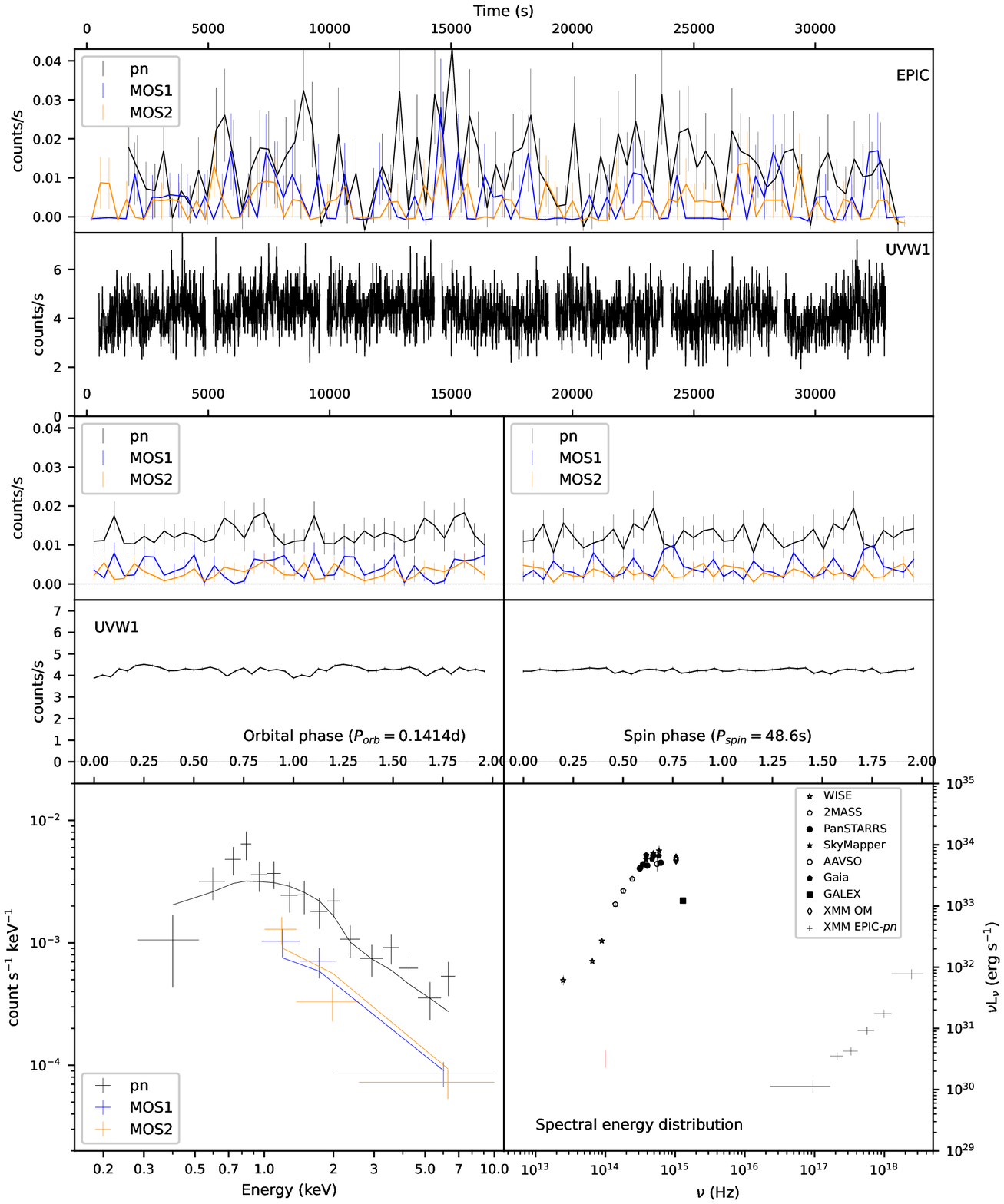}
    \caption{{\it Top panels:} X-ray and UV light curves of GI~Mon; {\it Middle panels:} phase folded on orbital and spin periods; {\it Bottom panels:} X-ray spectrum and SED.}
    \label{fig:gimon_fullpage}
\end{figure*}

\begin{figure*}[h!]
    \includegraphics{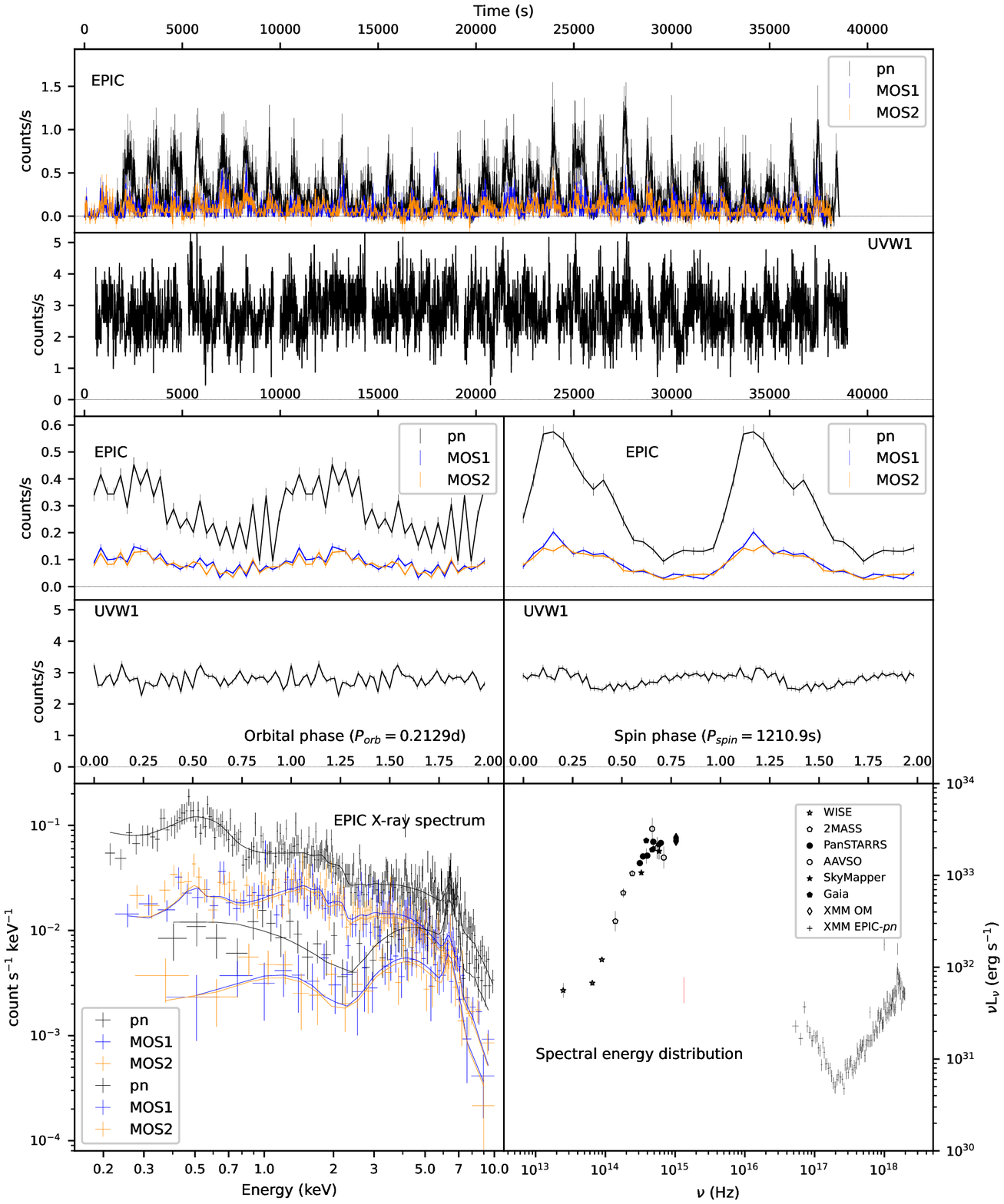}
    \caption{{\it Top panels:} X-ray and UV light curves of HZ~Pup; {\it Middle panels:} phase folded on orbital and spin periods; {\it Bottom panels:} X-ray spectrum and SED.}
    \label{fig:hzpup_fullpage}
\end{figure*}

\begin{figure*}[h!]
    \includegraphics{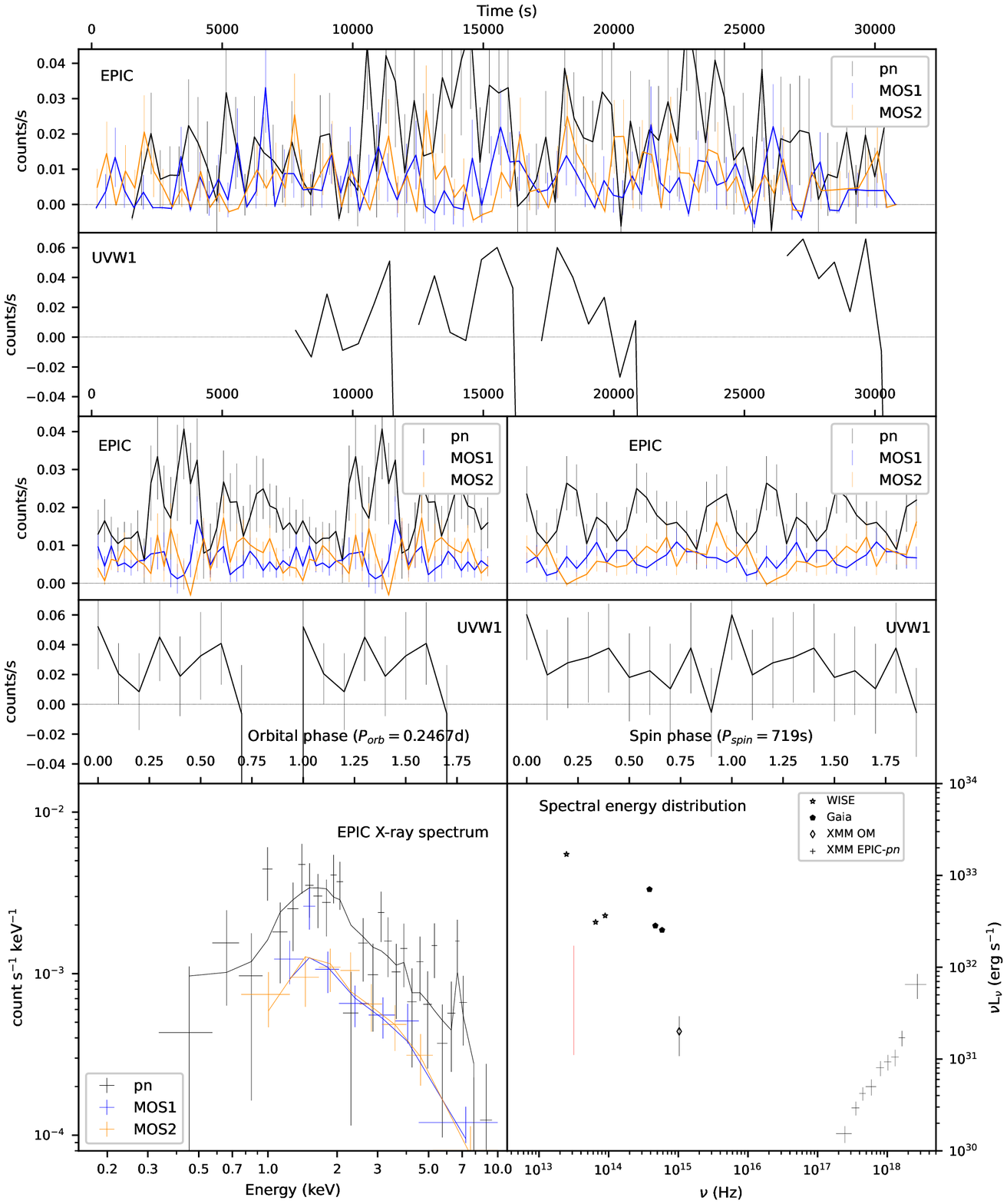}
    \caption{{\it Top panels:} X-ray and UV light curves of V1039~Cen; {\it Middle panels:} phase folded on the orbital and spin periods found in \cite{WoudtEtAl2005};{\it Bottom panels:} X-ray spectrum and SED.}
    \label{fig:v1039cen_fullpage}
\end{figure*}

\begin{figure*}[h!]
    \includegraphics{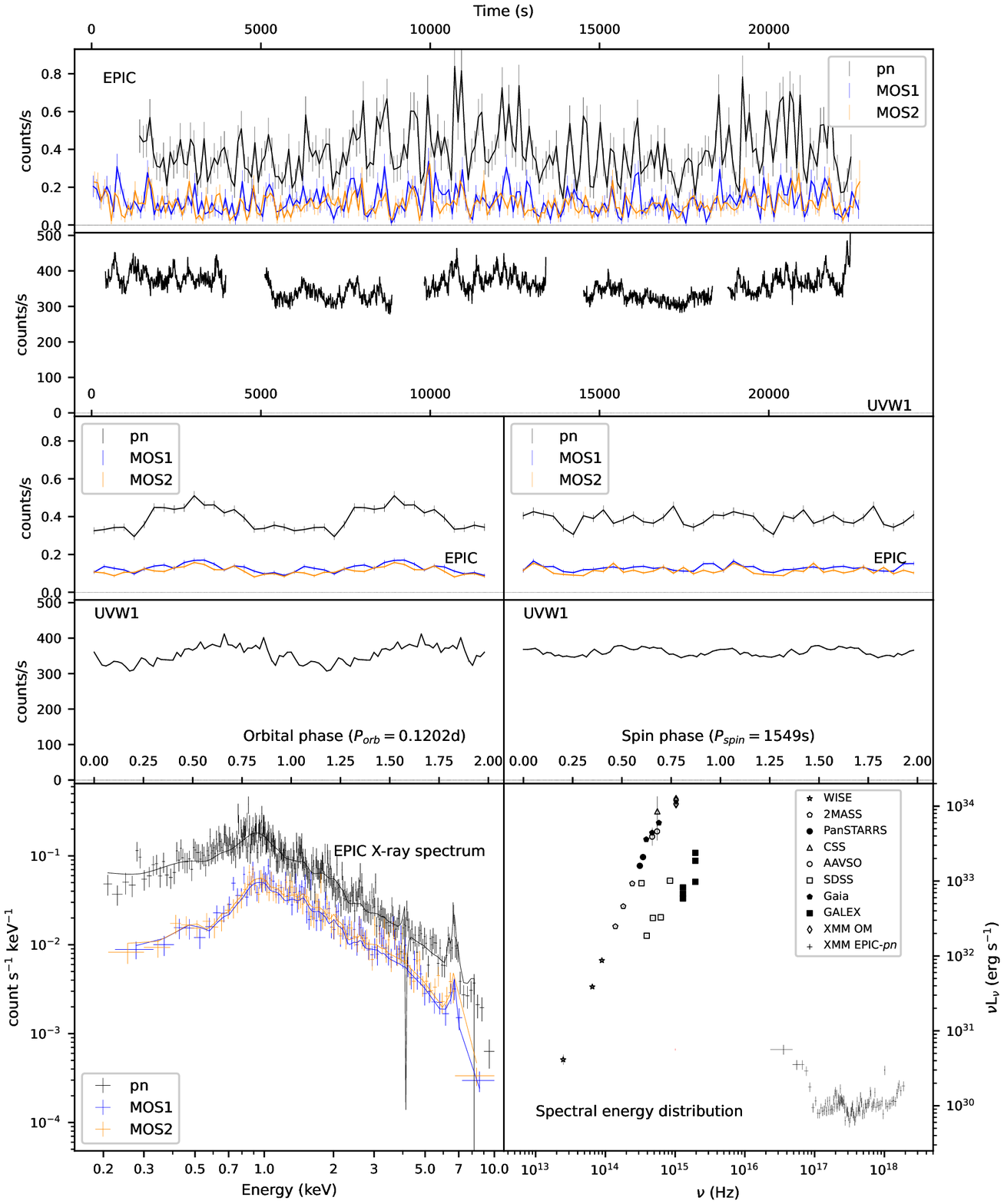}
    \caption{{\it Top panels:} X-ray and UV light curves of V1084 Her; {\it Middle panels:} phase folded on orbital and spin periods; {\it Bottom panels:} X-ray spectrum and SED.}
    \label{fig:v1084her_fullpage}
\end{figure*}

\begin{figure*}[h!]
    \includegraphics{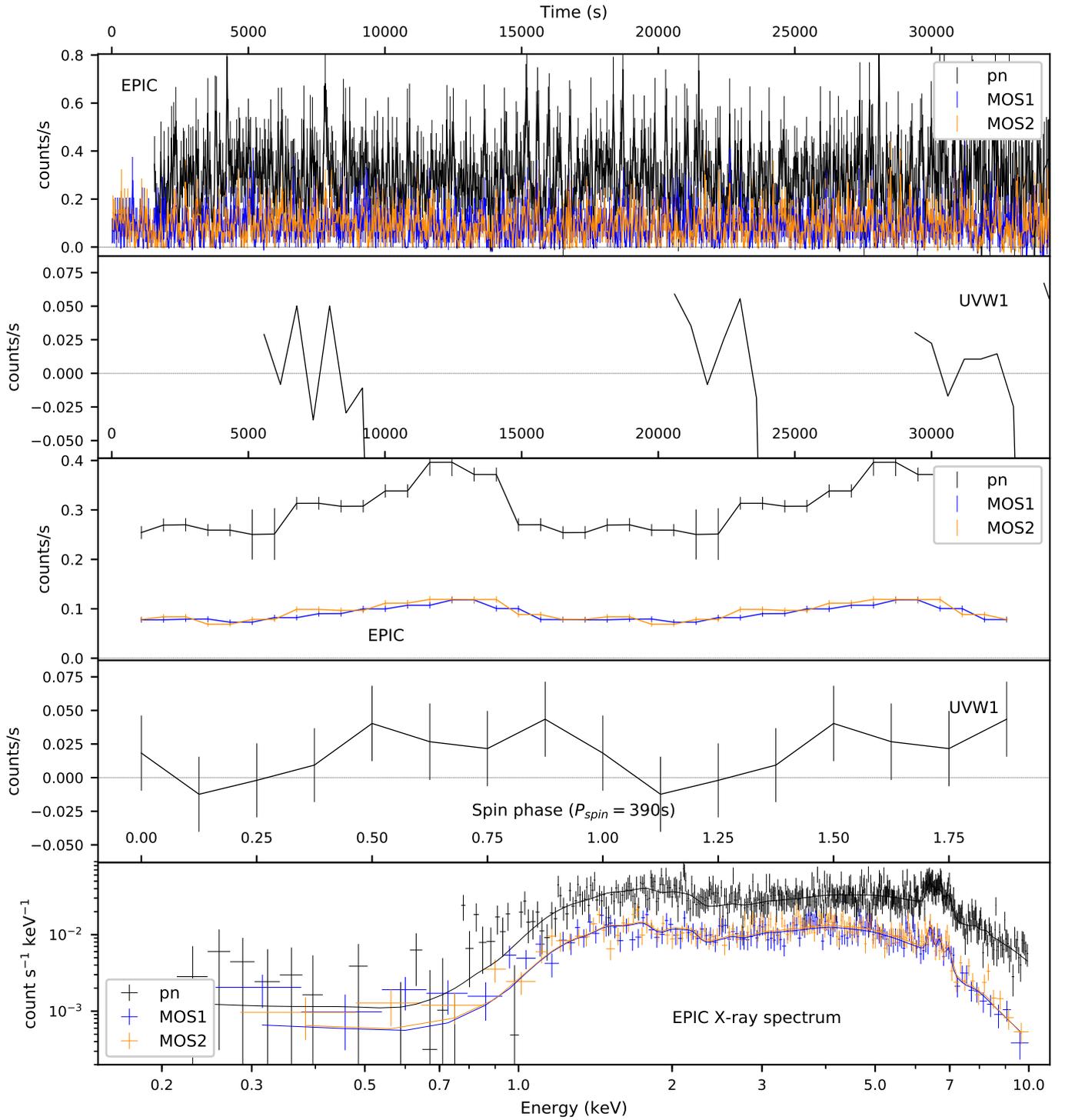}
    \caption{{\it Top panels:} X-ray and UV light curves of IGR J18151-1052; {\it Middle panels:} phase folded on spin period; {\it Bottom panel:} X-ray spectrum .}
    \label{fig:igr_fullpage}
\end{figure*}

\begin{figure*}[h!]
    \includegraphics{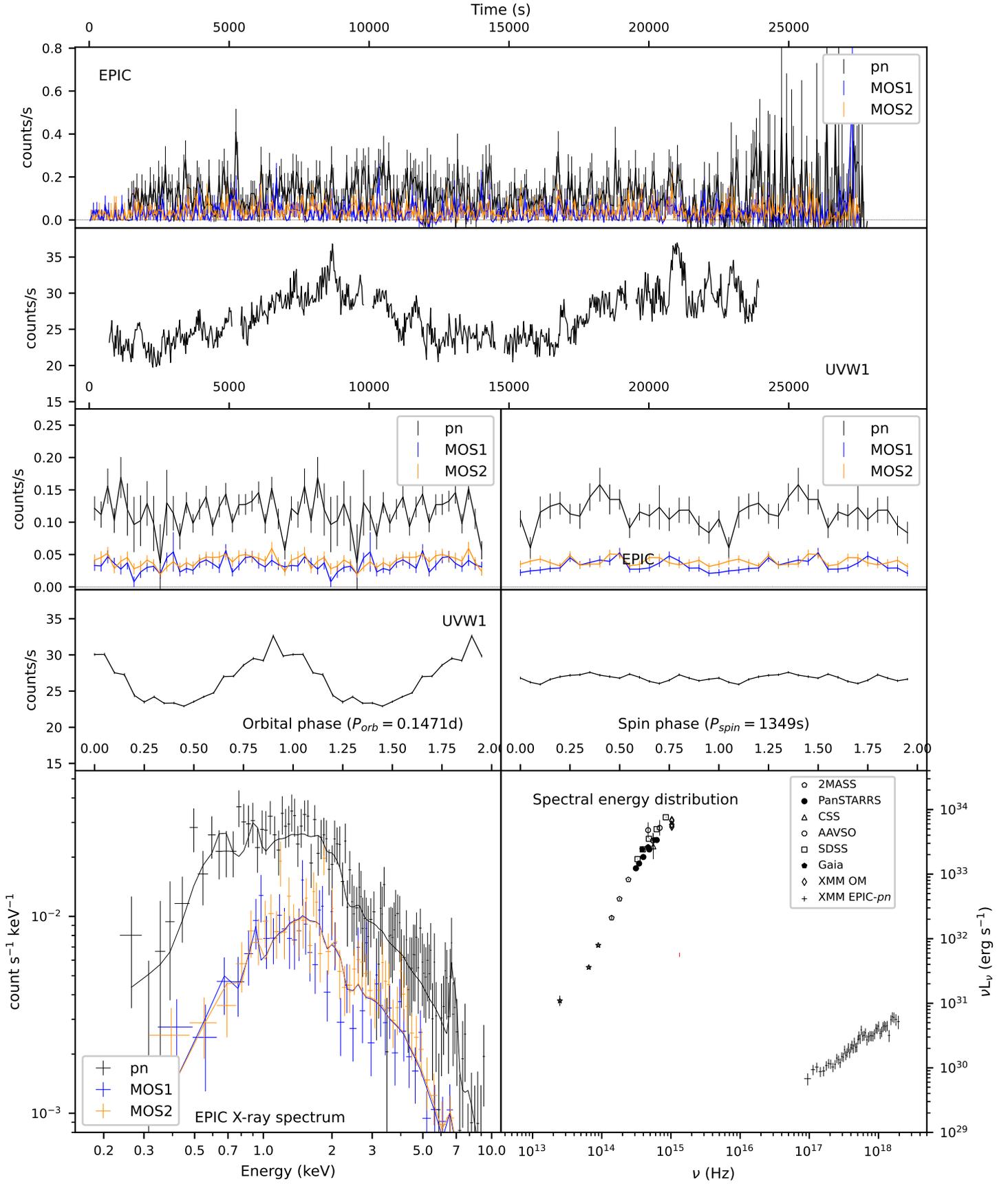}
    \caption{{\it Top panels:} X-ray and UV light curves of V533 Her; {\it Middle panels:} phase folded on spin period; {\it Bottom panel:} X-ray spectrum.}
    \label{fig:V533Her_fullpage}
\end{figure*}

\begin{figure*}[h!]
    \includegraphics{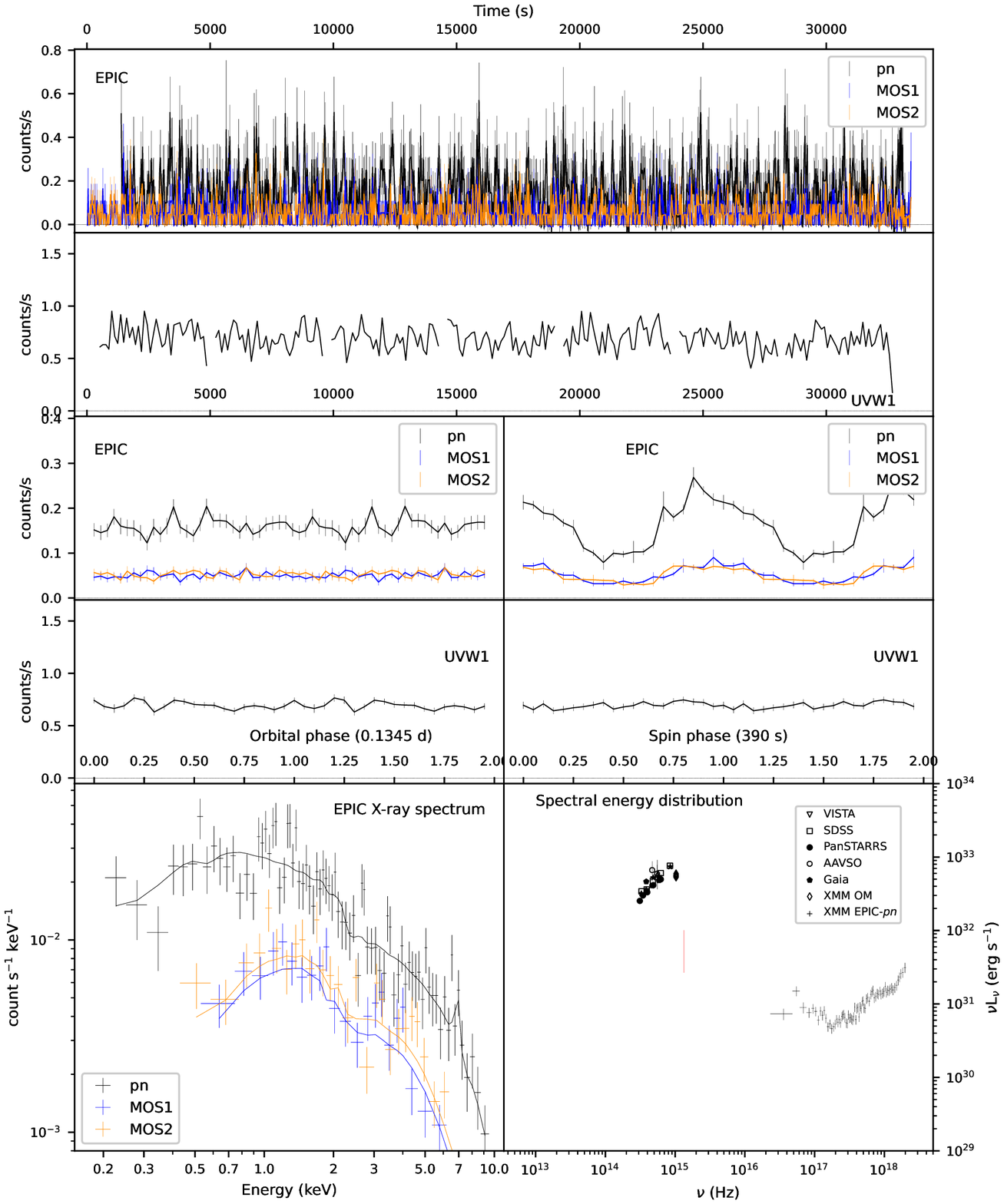}
    \caption{{\it Top panels:} X-ray and UV light curves of V349 Aqr; {\it Middle panels:} phase folded on orbital and spin periods; {\it Bottom panels:} X-ray spectrum and SED .}
    \label{fig:v349aqr_fullpage}
\end{figure*}

\appendix

\bibliography{bibli}

\begin{thebibliography}{98}
\expandafter\ifx\csname natexlab\endcsname\relax\def\natexlab#1{#1}\fi

\bibitem[{{Abbott} \& {Shafter}(1997)}]{AbbottShafter1997}
{Abbott}, T.~M.~C. \& {Shafter}, A.~W. 1997, in Astronomical Society of the
  Pacific Conference Series, Vol. 121, IAU Colloq. 163: Accretion Phenomena and
  Related Outflows, ed. D.~T. {Wickramasinghe}, G.~V. {Bicknell}, \&
  L.~{Ferrario}, 679

\bibitem[{{Adams} \& {Joy}(1918)}]{AdamsJoy1918}
{Adams}, W.~S. \& {Joy}, A.~H. 1918, \pasp, 30, 162

\bibitem[{{Anders} \& {Grevesse}(1989)}]{AndersGrevesse1989}
{Anders}, E. \& {Grevesse}, N. 1989, \gca, 53, 197

\bibitem[{{Arnaud}(1996)}]{Arnaud1996}
{Arnaud}, K.~A. 1996, in {Astronomical Society of the Pacific Conference
  Series}, Vol. 101, {Astronomical Data Analysis Software and Systems V}, ed.
  {G.~H.~Jacoby \& J.~Barnes}, 17

\bibitem[{{Astropy Collaboration} {et~al.}(2018){Astropy Collaboration},
  {Price-Whelan}, {Sip{\H{o}}cz}, {G{\"u}nther}, {Lim}, {Crawford}, {Conseil},
  {Shupe}, {Craig}, {Dencheva}, {Ginsburg}, {Vand erPlas}, {Bradley},
  {P{\'e}rez-Su{\'a}rez}, {de Val-Borro}, {Aldcroft}, {Cruz}, {Robitaille},
  {Tollerud}, {Ardelean}, {Babej}, {Bach}, {Bachetti}, {Bakanov}, {Bamford},
  {Barentsen}, {Barmby}, {Baumbach}, {Berry}, {Biscani}, {Boquien}, {Bostroem},
  {Bouma}, {Brammer}, {Bray}, {Breytenbach}, {Buddelmeijer}, {Burke},
  {Calderone}, {Cano Rodr{\'\i}guez}, {Cara}, {Cardoso}, {Cheedella}, {Copin},
  {Corrales}, {Crichton}, {D'Avella}, {Deil}, {Depagne}, {Dietrich}, {Donath},
  {Droettboom}, {Earl}, {Erben}, {Fabbro}, {Ferreira}, {Finethy}, {Fox},
  {Garrison}, {Gibbons}, {Goldstein}, {Gommers}, {Greco}, {Greenfield},
  {Groener}, {Grollier}, {Hagen}, {Hirst}, {Homeier}, {Horton}, {Hosseinzadeh},
  {Hu}, {Hunkeler}, {Ivezi{\'c}}, {Jain}, {Jenness}, {Kanarek}, {Kendrew},
  {Kern}, {Kerzendorf}, {Khvalko}, {King}, {Kirkby}, {Kulkarni}, {Kumar},
  {Lee}, {Lenz}, {Littlefair}, {Ma}, {Macleod}, {Mastropietro}, {McCully},
  {Montagnac}, {Morris}, {Mueller}, {Mumford}, {Muna}, {Murphy}, {Nelson},
  {Nguyen}, {Ninan}, {N{\"o}the}, {Ogaz}, {Oh}, {Parejko}, {Parley}, {Pascual},
  {Patil}, {Patil}, {Plunkett}, {Prochaska}, {Rastogi}, {Reddy Janga},
  {Sabater}, {Sakurikar}, {Seifert}, {Sherbert}, {Sherwood-Taylor}, {Shih},
  {Sick}, {Silbiger}, {Singanamalla}, {Singer}, {Sladen}, {Sooley},
  {Sornarajah}, {Streicher}, {Teuben}, {Thomas}, {Tremblay}, {Turner},
  {Terr{\'o}n}, {van Kerkwijk}, {de la Vega}, {Watkins}, {Weaver}, {Whitmore},
  {Woillez}, {Zabalza}, \& {Astropy Contributors}}]{Astropy2018}
{Astropy Collaboration}, {Price-Whelan}, A.~M., {Sip{\H{o}}cz}, B.~M., {et~al.}
  2018, \aj, 156, 123

\bibitem[{{Astropy Collaboration} {et~al.}(2013){Astropy Collaboration},
  {Robitaille}, {Tollerud}, {Greenfield}, {Droettboom}, {Bray}, {Aldcroft},
  {Davis}, {Ginsburg}, {Price-Whelan}, {Kerzendorf}, {Conley}, {Crighton},
  {Barbary}, {Muna}, {Ferguson}, {Grollier}, {Parikh}, {Nair}, {Unther},
  {Deil}, {Woillez}, {Conseil}, {Kramer}, {Turner}, {Singer}, {Fox}, {Weaver},
  {Zabalza}, {Edwards}, {Azalee Bostroem}, {Burke}, {Casey}, {Crawford},
  {Dencheva}, {Ely}, {Jenness}, {Labrie}, {Lim}, {Pierfederici}, {Pontzen},
  {Ptak}, {Refsdal}, {Servillat}, \& {Streicher}}]{Astropy2013}
{Astropy Collaboration}, {Robitaille}, T.~P., {Tollerud}, E.~J., {et~al.} 2013,
  \aap, 558, A33

\bibitem[{{Aungwerojwit} {et~al.}(2012){Aungwerojwit}, {G{\"a}nsicke},
  {Wheatley}, {Pyrzas}, {Staels}, {Krajci}, \&
  {Rodr{\'{\i}}guez-Gil}}]{AungwerojwitEtAl2012}
{Aungwerojwit}, A., {G{\"a}nsicke}, B.~T., {Wheatley}, P.~J., {et~al.} 2012,
  \apj, 758, 79

\bibitem[{{Bailer-Jones} {et~al.}(2018){Bailer-Jones}, {Rybizki}, {Fouesneau},
  {Mantelet}, \& {Andrae}}]{Bailer-JonesEtAl2018}
{Bailer-Jones}, C.~A.~L., {Rybizki}, J., {Fouesneau}, M., {Mantelet}, G., \&
  {Andrae}, R. 2018, \aj, 156, 58

\bibitem[{{Berg} {et~al.}(1992){Berg}, {Wegner}, {Foltz}, {Chaffee}, \&
  {Hewett}}]{BergEtAl1992}
{Berg}, C., {Wegner}, G., {Foltz}, C.~B., {Chaffee}, Jr., F.~H., \& {Hewett},
  P.~C. 1992, \apjs, 78, 409

\bibitem[{{Blackburn}(1995)}]{Blackburn1995}
{Blackburn}, J.~K. 1995, in {Astronomical Society of the Pacific Conference
  Series}, Vol.~77, {Astronomical Data Analysis Software and Systems IV}, ed.
  R.~A. {Shaw}, H.~E. {Payne}, \& J.~J.~E. {Hayes}, 367

\bibitem[{{Blanton} {et~al.}(2017){Blanton}, {Bershady}, {Abolfathi},
  {Albareti}, {Allende Prieto}, {Almeida}, {Alonso-Garc{\'{\i}}a}, {Anders},
  {Anderson}, {Andrews}, \& et~al.}]{BlantonEtAl2017}
{Blanton}, M.~R., {Bershady}, M.~A., {Abolfathi}, B., {et~al.} 2017, \aj, 154,
  28

\bibitem[{{Brown} {et~al.}(2003){Brown}, {Yamamoto}, {Nakano}, {Kushida},
  {Kushida}, {Kadota}, {Gilmore}, {Kilmartin}, {Skuljan}, {Stubbings},
  {Waagen}, \& {Jones}}]{BrownEtAl2003}
{Brown}, N.~J., {Yamamoto}, M., {Nakano}, S., {et~al.} 2003, \iaucirc, 8123, 1

\bibitem[{{Brunschweiger} {et~al.}(2009){Brunschweiger}, {Greiner}, {Ajello},
  \& {Osborne}}]{BrunschweigerEtAl2009}
{Brunschweiger}, J., {Greiner}, J., {Ajello}, M., \& {Osborne}, J. 2009, \aap,
  496, 121

\bibitem[{{Burrows} {et~al.}(2005){Burrows}, {Hill}, {Nousek}, {Kennea},
  {Wells}, {Osborne}, {Abbey}, {Beardmore}, {Mukerjee}, {Short}, {Chincarini},
  {Campana}, {Citterio}, {Moretti}, {Pagani}, {Tagliaferri}, {Giommi},
  {Capalbi}, {Tamburelli}, {Angelini}, {Cusumano}, {Br{\"a}uninger}, {Burkert},
  \& {Hartner}}]{BurrowsEtAl2005}
{Burrows}, D.~N., {Hill}, J.~E., {Nousek}, J.~A., {et~al.} 2005, \ssr, 120, 165

\bibitem[{{Cash}(1979)}]{Cash1979}
{Cash}, W. 1979, \apj, 228, 939

\bibitem[{{Chambers} {et~al.}(2016){Chambers}, {Magnier}, {Metcalfe},
  {Flewelling}, {Huber}, {Waters}, {Denneau}, {Draper}, {Farrow}, {Finkbeiner},
  {Holmberg}, {Koppenhoefer}, {Price}, {Saglia}, {Schlafly}, {Smartt},
  {Sweeney}, {Wainscoat}, {Burgett}, {Grav}, {Heasley}, {Hodapp}, {Jedicke},
  {Kaiser}, {Kudritzki}, {Luppino}, {Lupton}, {Monet}, {Morgan}, {Onaka},
  {Stubbs}, {Tonry}, {Banados}, {Bell}, {Bender}, {Bernard}, {Botticella},
  {Casertano}, {Chastel}, {Chen}, {Chen}, {Cole}, {Deacon}, {Frenk},
  {Fitzsimmons}, {Gezari}, {Goessl}, {Goggia}, {Goldman}, {Grebel}, {Hambly},
  {Hasinger}, {Heavens}, {Heckman}, {Henderson}, {Henning}, {Holman}, {Hopp},
  {Ip}, {Isani}, {Keyes}, {Koekemoer}, {Kotak}, {Long}, {Lucey}, {Liu},
  {Martin}, {McLean}, {Morganson}, {Murphy}, {Nieto-Santisteban}, {Norberg},
  {Peacock}, {Pier}, {Postman}, {Primak}, {Rae}, {Rest}, {Riess}, {Riffeser},
  {Rix}, {Roser}, {Schilbach}, {Schultz}, {Scolnic}, {Szalay}, {Seitz},
  {Shiao}, {Small}, {Smith}, {Soderblom}, {Taylor}, {Thakar}, {Thiel},
  {Thilker}, {Urata}, {Valenti}, {Walter}, {Watters}, {Werner}, {White},
  {Wood-Vasey}, \& {Wyse}}]{ChambersEtAl2016}
{Chambers}, K.~C., {Magnier}, E.~A., {Metcalfe}, N., {et~al.} 2016, ArXiv
  e-prints [\eprint[arXiv]{1612.05560}]

\bibitem[{{Cohen} {et~al.}(2003){Cohen}, {Wheaton}, \&
  {Megeath}}]{CohenEtAl2003}
{Cohen}, M., {Wheaton}, W.~A., \& {Megeath}, S.~T. 2003, \aj, 126, 1090

\bibitem[{{Cross} {et~al.}(2012){Cross}, {Collins}, {Mann}, {Read}, {Sutorius},
  {Blake}, {Holliman}, {Hambly}, {Emerson}, {Lawrence}, \&
  {Noddle}}]{CrossEtAl2012}
{Cross}, N.~J.~G., {Collins}, R.~S., {Mann}, R.~G., {et~al.} 2012, \aap, 548,
  A119

\bibitem[{{Cs{\'a}k} {et~al.}(2005){Cs{\'a}k}, {Kiss}, {Retter}, {Jacob}, \&
  {Kaspi}}]{CsakEtAl2005}
{Cs{\'a}k}, B., {Kiss}, L.~L., {Retter}, A., {Jacob}, A., \& {Kaspi}, S. 2005,
  \aap, 429, 599

\bibitem[{{Cusumano} {et~al.}(2010){Cusumano}, {La Parola}, {Segreto},
  {Ferrigno}, {Maselli}, {Sbarufatti}, {Romano}, {Chincarini}, {Giommi},
  {Masetti}, {Moretti}, {Parisi}, \& {Tagliaferri}}]{CusumanoEtAl2010}
{Cusumano}, G., {La Parola}, V., {Segreto}, A., {et~al.} 2010, \aap, 524, A64

\bibitem[{{de Jager} {et~al.}(1989){de Jager}, {Raubenheimer}, \&
  {Swanepoel}}]{deJagerEtAl1989}
{de Jager}, O.~C., {Raubenheimer}, B.~C., \& {Swanepoel}, J.~W.~H. 1989, \aap,
  221, 180

\bibitem[{{den Herder} {et~al.}(2001){den Herder}, {Brinkman}, {Kahn},
  {Branduardi-Raymont}, {Thomsen}, {Aarts}, {Audard}, {Bixler}, {den Boggende},
  {Cottam}, {Decker}, {Dubbeldam}, {Erd}, {Goulooze}, {G{\"u}del}, {Guttridge},
  {Hailey}, {Janabi}, {Kaastra}, {de Korte}, {van Leeuwen}, {Mauche},
  {McCalden}, {Mewe}, {Naber}, {Paerels}, {Peterson}, {Rasmussen}, {Rees},
  {Sakelliou}, {Sako}, {Spodek}, {Stern}, {Tamura}, {Tandy}, {de Vries},
  {Welch}, \& {Zehnder}}]{denHerderEtAl2001}
{den Herder}, J.~W., {Brinkman}, A.~C., {Kahn}, S.~M., {et~al.} 2001, \aap,
  365, L7

\bibitem[{{Dobrotka} {et~al.}(2006){Dobrotka}, {Retter}, \&
  {Liu}}]{DobrotkaEtAl2006}
{Dobrotka}, A., {Retter}, A., \& {Liu}, A. 2006, \mnras, 371, 459

\bibitem[{{Drake} {et~al.}(2009){Drake}, {Djorgovski}, {Mahabal}, {Beshore},
  {Larson}, {Graham}, {Williams}, {Christensen}, {Catelan}, {Boattini},
  {Gibbs}, {Hill}, \& {Kowalski}}]{DrakeEtAl2009}
{Drake}, A.~J., {Djorgovski}, S.~G., {Mahabal}, A., {et~al.} 2009, \apj, 696,
  870

\bibitem[{{Ebisawa} {et~al.}(2008){Ebisawa}, {Yamauchi}, {Tanaka}, {Koyama},
  {Ezoe}, {Bamba}, {Kokubun}, {Hyodo}, {Tsujimoto}, \&
  {Takahashi}}]{EbisawaEtAl2008}
{Ebisawa}, K., {Yamauchi}, S., {Tanaka}, Y., {et~al.} 2008, \pasj, 60, S223

\bibitem[{Efron(1979)}]{Efron1979}
Efron, B. 1979, Ann. Statist., 7, 1

\bibitem[{{Ezuka} \& {Ishida}(1999)}]{EzukaIshida1999}
{Ezuka}, H. \& {Ishida}, M. 1999, \apjs, 120, 277

\bibitem[{{Fukugita} {et~al.}(1996){Fukugita}, {Ichikawa}, {Gunn}, {Doi},
  {Shimasaku}, \& {Schneider}}]{FukugitaEtAl1996}
{Fukugita}, M., {Ichikawa}, T., {Gunn}, J.~E., {et~al.} 1996, \aj, 111, 1748

\bibitem[{{Gaia Collaboration} {et~al.}(2018{\natexlab{a}}){Gaia
  Collaboration}, {Brown}, {Vallenari}, {Prusti}, {de Bruijne}, {Babusiaux}, \&
  {Bailer-Jones}}]{Gaia2018}
{Gaia Collaboration}, {Brown}, A.~G.~A., {Vallenari}, A., {et~al.}
  2018{\natexlab{a}}, ArXiv e-prints [\eprint[arXiv]{1804.09365}]

\bibitem[{{Gaia Collaboration} {et~al.}(2018{\natexlab{b}}){Gaia
  Collaboration}, {Brown}, {Vallenari}, {Prusti}, {de Bruijne}, {Babusiaux},
  {Bailer-Jones}, {Biermann}, {Evans}, {Eyer}, {Jansen}, {Jordi}, {Klioner},
  {Lammers}, {Lindegren}, {Luri}, {Mignard}, {Panem}, {Pourbaix}, {Randich},
  {Sartoretti}, {Siddiqui}, {Soubiran}, {van Leeuwen}, {Walton}, {Arenou},
  {Bastian}, {Cropper}, {Drimmel}, {Katz}, {Lattanzi}, {Bakker}, {Cacciari},
  {Casta{\~n}eda}, {Chaoul}, {Cheek}, {De Angeli}, {Fabricius}, {Guerra},
  {Holl}, {Masana}, {Messineo}, {Mowlavi}, {Nienartowicz}, {Panuzzo},
  {Portell}, {Riello}, {Seabroke}, {Tanga}, {Th{\'e}venin}, {Gracia-Abril},
  {Comoretto}, {Garcia-Reinaldos}, {Teyssier}, {Altmann}, {Andrae}, {Audard},
  {Bellas-Velidis}, {Benson}, {Berthier}, {Blomme}, {Burgess}, {Busso},
  {Carry}, {Cellino}, {Clementini}, {Clotet}, {Creevey}, {Davidson}, {De
  Ridder}, {Delchambre}, {Dell'Oro}, {Ducourant},
  {Fern{\'a}ndez-Hern{\'a}ndez}, {Fouesneau}, {Fr{\'e}mat}, {Galluccio},
  {Garc{\'\i}a-Torres}, {Gonz{\'a}lez-N{\'u}{\~n}ez}, {Gonz{\'a}lez-Vidal},
  {Gosset}, {Guy}, {Halbwachs}, {Hambly}, {Harrison}, {Hern{\'a}ndez},
  {Hestroffer}, {Hodgkin}, {Hutton}, {Jasniewicz}, {Jean-Antoine-Piccolo},
  {Jordan}, {Korn}, {Krone-Martins}, {Lanzafame}, {Lebzelter}, {L{\"o}ffler},
  {Manteiga}, {Marrese}, {Mart{\'\i}n-Fleitas}, {Moitinho}, {Mora}, {Muinonen},
  {Osinde}, {Pancino}, {Pauwels}, {Petit}, {Recio-Blanco}, {Richards},
  {Rimoldini}, {Robin}, {Sarro}, {Siopis}, {Smith}, {Sozzetti}, {S{\"u}veges},
  {Torra}, {van Reeven}, {Abbas}, {Abreu Aramburu}, {Accart}, {Aerts},
  {Altavilla}, {{\'A}lvarez}, {Alvarez}, {Alves}, {Anderson}, {Andrei},
  {Anglada Varela}, {Antiche}, {Antoja}, {Arcay}, {Astraatmadja}, {Bach},
  {Baker}, {Balaguer-N{\'u}{\~n}ez}, {Balm}, {Barache}, {Barata}, {Barbato},
  {Barblan}, {Barklem}, {Barrado}, {Barros}, {Barstow}, {Bartholom{\'e}
  Mu{\~n}oz}, {Bassilana}, {Becciani}, {Bellazzini}, {Berihuete}, {Bertone},
  {Bianchi}, {Bienaym{\'e}}, {Blanco-Cuaresma}, {Boch}, {Boeche}, {Bombrun},
  {Borrachero}, {Bossini}, {Bouquillon}, {Bourda}, {Bragaglia}, {Bramante},
  {Breddels}, {Bressan}, {Brouillet}, {Br{\"u}semeister}, {Brugaletta},
  {Bucciarelli}, {Burlacu}, {Busonero}, {Butkevich}, {Buzzi}, {Caffau},
  {Cancelliere}, {Cannizzaro}, {Cantat-Gaudin}, {Carballo}, {Carlucci},
  {Carrasco}, {Casamiquela}, {Castellani}, {Castro-Ginard}, {Charlot},
  {Chemin}, {Chiavassa}, {Cocozza}, {Costigan}, {Cowell}, {Crifo}, {Crosta},
  {Crowley}, {Cuypers}, {Dafonte}, {Damerdji}, {Dapergolas}, {David}, {David},
  {de Laverny}, {De Luise}, {De March}, {de Martino}, {de Souza}, {de Torres},
  {Debosscher}, {del Pozo}, {Delbo}, {Delgado}, {Delgado}, {Di Matteo},
  {Diakite}, {Diener}, {Distefano}, {Dolding}, {Drazinos}, {Dur{\'a}n},
  {Edvardsson}, {Enke}, {Eriksson}, {Esquej}, {Eynard Bontemps}, {Fabre},
  {Fabrizio}, {Faigler}, {Falc{\~a}o}, {Farr{\`a}s Casas}, {Federici},
  {Fedorets}, {Fernique}, {Figueras}, {Filippi}, {Findeisen}, {Fonti},
  {Fraile}, {Fraser}, {Fr{\'e}zouls}, {Gai}, {Galleti}, {Garabato},
  {Garc{\'\i}a-Sedano}, {Garofalo}, {Garralda}, {Gavel}, {Gavras}, {Gerssen},
  {Geyer}, {Giacobbe}, {Gilmore}, {Girona}, {Giuffrida}, {Glass}, {Gomes},
  {Granvik}, {Gueguen}, {Guerrier}, {Guiraud}, {Guti{\'e}rrez-S{\'a}nchez},
  {Haigron}, {Hatzidimitriou}, {Hauser}, {Haywood}, {Heiter}, {Helmi}, {Heu},
  {Hilger}, {Hobbs}, {Hofmann}, {Holland}, {Huckle}, {Hypki}, {Icardi},
  {Jan{\ss}en}, {Jevardat de Fombelle}, {Jonker}, {Juh{\'a}sz}, {Julbe},
  {Karampelas}, {Kewley}, {Klar}, {Kochoska}, {Kohley}, {Kolenberg},
  {Kontizas}, {Kontizas}, {Koposov}, {Kordopatis}, {Kostrzewa-Rutkowska},
  {Koubsky}, {Lambert}, {Lanza}, {Lasne}, {Lavigne}, {Le Fustec}, {Le
  Poncin-Lafitte}, {Lebreton}, {Leccia}, {Leclerc}, {Lecoeur-Taibi},
  {Lenhardt}, {Leroux}, {Liao}, {Licata}, {Lindstr{\o}m}, {Lister}, {Livanou},
  {Lobel}, {L{\'o}pez}, {Managau}, {Mann}, {Mantelet}, {Marchal}, {Marchant},
  {Marconi}, {Marinoni}, {Marschalk{\'o}}, {Marshall}, {Martino}, {Marton},
  {Mary}, {Massari}, {Matijevi{\v{c}}}, {Mazeh}, {McMillan}, {Messina},
  {Michalik}, {Millar}, {Molina}, {Molinaro}, {Moln{\'a}r}, {Montegriffo},
  {Mor}, {Morbidelli}, {Morel}, {Morris}, {Mulone}, {Muraveva}, {Musella},
  {Nelemans}, {Nicastro}, {Noval}, {O'Mullane}, {Ord{\'e}novic},
  {Ord{\'o}{\~n}ez-Blanco}, {Osborne}, {Pagani}, {Pagano}, {Pailler},
  {Palacin}, {Palaversa}, {Panahi}, {Pawlak}, {Piersimoni}, {Pineau}, {Plachy},
  {Plum}, {Poggio}, {Poujoulet}, {Pr{\v{s}}a}, {Pulone}, {Racero}, {Ragaini},
  {Rambaux}, {Ramos-Lerate}, {Regibo}, {Reyl{\'e}}, {Riclet}, {Ripepi}, {Riva},
  {Rivard}, {Rixon}, {Roegiers}, {Roelens}, {Romero-G{\'o}mez}, {Rowell},
  {Royer}, {Ruiz-Dern}, {Sadowski}, {Sagrist{\`a} Sell{\'e}s}, {Sahlmann},
  {Salgado}, {Salguero}, {Sanna}, {Santana-Ros}, {Sarasso}, {Savietto},
  {Schultheis}, {Sciacca}, {Segol}, {Segovia}, {S{\'e}gransan}, {Shih},
  {Siltala}, {Silva}, {Smart}, {Smith}, {Solano}, {Solitro}, {Sordo}, {Soria
  Nieto}, {Souchay}, {Spagna}, {Spoto}, {Stampa}, {Steele},
  {Steidelm{\"u}ller}, {Stephenson}, {Stoev}, {Suess}, {Surdej}, {Szabados},
  {Szegedi-Elek}, {Tapiador}, {Taris}, {Tauran}, {Taylor}, {Teixeira},
  {Terrett}, {Teyssand ier}, {Thuillot}, {Titarenko}, {Torra Clotet}, {Turon},
  {Ulla}, {Utrilla}, {Uzzi}, {Vaillant}, {Valentini}, {Valette}, {van Elteren},
  {Van Hemelryck}, {van Leeuwen}, {Vaschetto}, {Vecchiato}, {Veljanoski},
  {Viala}, {Vicente}, {Vogt}, {von Essen}, {Voss}, {Votruba}, {Voutsinas},
  {Walmsley}, {Weiler}, {Wertz}, {Wevers}, {Wyrzykowski}, {Yoldas},
  {{\v{Z}}erjal}, {Ziaeepour}, {Zorec}, {Zschocke}, {Zucker}, {Zurbach}, \&
  {Zwitter}}]{Gaia2016}
{Gaia Collaboration}, {Brown}, A.~G.~A., {Vallenari}, A., {et~al.}
  2018{\natexlab{b}}, \aap, 616, A1

\bibitem[{{Haberl} {et~al.}(2002){Haberl}, {Motch}, \&
  {Zickgraf}}]{HaberlEtAl2002}
{Haberl}, F., {Motch}, C., \& {Zickgraf}, F.-J. 2002, \aap, 387, 201

\bibitem[{{HI4PI Collaboration} {et~al.}(2016){HI4PI Collaboration}, {Ben
  Bekhti}, {Fl{\"o}er}, {Keller}, {Kerp}, {Lenz}, {Winkel}, {Bailin},
  {Calabretta}, {Dedes}, {Ford}, {Gibson}, {Haud}, {Janowiecki}, {Kalberla},
  {Lockman}, {McClure-Griffiths}, {Murphy}, {Nakanishi}, {Pisano}, \&
  {Staveley-Smith}}]{BekhtiEtAl2016}
{HI4PI Collaboration}, {Ben Bekhti}, N., {Fl{\"o}er}, L., {et~al.} 2016, \aap,
  594, A116

\bibitem[{{Hoffmeister}(1964)}]{Hoffmeister1964}
{Hoffmeister}, C. 1964, Information Bulletin on Variable Stars, 45, 1

\bibitem[{{Humphrey} {et~al.}(2009){Humphrey}, {Liu}, \&
  {Buote}}]{HumphreyEtAl2009}
{Humphrey}, P.~J., {Liu}, W., \& {Buote}, D.~A. 2009, \apj, 693, 822

\bibitem[{{Jarrett} {et~al.}(2011){Jarrett}, {Cohen}, {Masci}, {Wright},
  {Stern}, {Benford}, {Blain}, {Carey}, {Cutri}, {Eisenhardt}, {Lonsdale},
  {Mainzer}, {Marsh}, {Padgett}, {Petty}, {Ressler}, {Skrutskie}, {Stanford},
  {Surace}, {Tsai}, {Wheelock}, \& {Yan}}]{JarrettEtAl2011}
{Jarrett}, T.~H., {Cohen}, M., {Masci}, F., {et~al.} 2011, \apj, 735, 112

\bibitem[{{Kaastra}(2017)}]{Kaastra2017}
{Kaastra}, J.~S. 2017, \aap, 605, A51

\bibitem[{{Kirsch} {et~al.}(2004){Kirsch}, {Altieri}, {Chen}, {Haberl},
  {Metcalfe}, {Pollock}, {Read}, {Saxton}, {Sembay}, \&
  {Smith}}]{KirschEtAl2004}
{Kirsch}, M.~G.~F., {Altieri}, B., {Chen}, B., {et~al.} 2004, in Society of
  Photo-Optical Instrumentation Engineers (SPIE) Conference Series, Vol. 5488,
  UV and Gamma-Ray Space Telescope Systems, ed. G.~{Hasinger} \& M.~J.~L.
  {Turner}, 103--114

\bibitem[{{Kraft} {et~al.}(1991){Kraft}, {Burrows}, \&
  {Nousek}}]{KraftEtAl1991}
{Kraft}, R.~P., {Burrows}, D.~N., \& {Nousek}, J.~A. 1991, \apj, 374, 344

\bibitem[{{Krivonos} {et~al.}(2009){Krivonos}, {Tsygankov}, {Sunyaev},
  {Melnikov}, {Bikmaev}, {Pavlinsky}, \& {Burenin}}]{KrivonosEtAl2009}
{Krivonos}, R., {Tsygankov}, S., {Sunyaev}, R., {et~al.} 2009, The Astronomer's
  Telegram, 2170, 1

\bibitem[{{Liller} {et~al.}(2001){Liller}, {Gilmore}, {Jones}, {Pearce},
  {Monard}, {Africa}, \& {Amorim}}]{LillerEtAl2001}
{Liller}, W., {Gilmore}, A.~C., {Jones}, A., {et~al.} 2001, \iaucirc, 7726, 1

\bibitem[{{Liu} \& {Hu}(2000)}]{LiuHu2000}
{Liu}, W. \& {Hu}, J.~Y. 2000, \apjs, 128, 387

\bibitem[{{Loredo}(1992)}]{Loredo1992}
{Loredo}, T.~J. 1992, in Statistical Challenges in Modern Astronomy, ed. E.~D.
  {Feigelson} \& G.~J. {Babu}, 275--297

\bibitem[{{Martin} {et~al.}(2005){Martin}, {Fanson}, {Schiminovich},
  {Morrissey}, {Friedman}, {Barlow}, {Conrow}, {Grange}, {Jelinsky},
  {Milliard}, {Siegmund}, {Bianchi}, {Byun}, {Donas}, {Forster}, {Heckman},
  {Lee}, {Madore}, {Malina}, {Neff}, {Rich}, {Small}, {Surber}, {Szalay},
  {Welsh}, \& {Wyder}}]{MartinEtAl2005}
{Martin}, D.~C., {Fanson}, J., {Schiminovich}, D., {et~al.} 2005, \apjl, 619,
  L1

\bibitem[{{Masetti} {et~al.}(2013){Masetti}, {Parisi}, {Palazzi},
  {Jim{\'e}nez-Bail{\'o}n}, {Chavushyan}, {McBride}, {Rojas}, {Steward},
  {Bassani}, {Bazzano}, {Bird}, {Charles}, {Galaz}, {Landi}, {Malizia},
  {Mason}, {Minniti}, {Morelli}, {Schiavone}, {Stephen}, \&
  {Ubertini}}]{MasettiEtAl2013}
{Masetti}, N., {Parisi}, P., {Palazzi}, E., {et~al.} 2013, \aap, 556, A120

\bibitem[{{Mason} {et~al.}(2001){Mason}, {Breeveld}, {Much}, {Carter},
  {Cordova}, {Cropper}, {Fordham}, {Huckle}, {Ho}, {Kawakami}, {Kennea},
  {Kennedy}, {Mittaz}, {Pandel}, {Priedhorsky}, {Sasseen}, {Shirey}, {Smith},
  \& {Vreux}}]{MasonEtAl2001}
{Mason}, K.~O., {Breeveld}, A., {Much}, R., {et~al.} 2001, \aap, 365, L36

\bibitem[{{Mason} {et~al.}(1992){Mason}, {Watson}, {Ponman}, {Charles}, {Duck},
  {Hassall}, {Howell}, {Ishida}, {Jones}, \& {Mittaz}}]{MasonEtAl1992}
{Mason}, K.~O., {Watson}, M.~G., {Ponman}, T.~J., {et~al.} 1992, \mnras, 258,
  749

\bibitem[{{Mewe} {et~al.}(1985){Mewe}, {Gronenschild}, \& {van den
  Oord}}]{MeweEtAl1985}
{Mewe}, R., {Gronenschild}, E.~H.~B.~M., \& {van den Oord}, G.~H.~J. 1985,
  \aaps, 62, 197

\bibitem[{{Mickaelian} {et~al.}(2002){Mickaelian}, {Balayan}, {Ilovaisky},
  {Chevalier}, {V{\'e}ron-Cetty}, \& {V{\'e}ron}}]{MickaelianEtAl2002}
{Mickaelian}, A.~M., {Balayan}, S.~K., {Ilovaisky}, S.~A., {et~al.} 2002, \aap,
  381, 894

\bibitem[{{Morrissey} {et~al.}(2007){Morrissey}, {Conrow}, {Barlow}, {Small},
  {Seibert}, {Wyder}, {Budav{\'a}ri}, {Arnouts}, {Friedman}, {Forster},
  {Martin}, {Neff}, {Schiminovich}, {Bianchi}, {Donas}, {Heckman}, {Lee},
  {Madore}, {Milliard}, {Rich}, {Szalay}, {Welsh}, \& {Yi}}]{MorrisseyEtAl2007}
{Morrissey}, P., {Conrow}, T., {Barlow}, T.~A., {et~al.} 2007, \apjs, 173, 682

\bibitem[{{Mukai}(2017)}]{Mukai2017}
{Mukai}, K. 2017, \pasp, 129, 062001

\bibitem[{{Mukai} {et~al.}(2003){Mukai}, {Still}, \&
  {Ringwald}}]{MukaiEtAl2003}
{Mukai}, K., {Still}, M., \& {Ringwald}, F.~A. 2003, \apj, 594, 428

\bibitem[{{Nakano} {et~al.}(1995){Nakano}, {Takamizawa}, {Kushida}, {Waagen},
  {Scovil}, \& {Collins}}]{NakanoEtAl1995}
{Nakano}, S., {Takamizawa}, K., {Kushida}, Y., {et~al.} 1995, \iaucirc, 6133, 1

\bibitem[{{Ness} {et~al.}(2008){Ness}, {Starrfield}, {Schwarz}, {Osborne},
  {Page}, {Rudy}, {Russell}, {Lynch}, {Woodward}, \& {Rivkin}}]{NessEtAl2008}
{Ness}, J.~U., {Starrfield}, S., {Schwarz}, G.~J., {et~al.} 2008, International
  Astronomical Union Circular, 8911, 2

\bibitem[{{Norton} \& {Watson}(1989)}]{NortonWatson1989}
{Norton}, A.~J. \& {Watson}, M.~G. 1989, \mnras, 237, 853

\bibitem[{{Nucita} {et~al.}(2019){Nucita}, {Conversi}, \&
  {Licchelli}}]{NucitaEtAl2019}
{Nucita}, A.~A., {Conversi}, L., \& {Licchelli}, D. 2019, \mnras, 484, 3119

\bibitem[{{Page} {et~al.}(2014){Page}, {Yershov}, {Breeveld}, {Kuin},
  {Mignani}, {Smith}, {Rawlings}, {Oates}, {Siegel}, \&
  {Roming}}]{PageEtAl2014}
{Page}, M.~J., {Yershov}, V., {Breeveld}, A., {et~al.} 2014, in Proceedings of
  Swift: 10 Years of Discovery (SWIFT 10), held 2-5 December 2014 at La
  Sapienza University, Rome, Italy., 37

\bibitem[{{Parker} {et~al.}(2005){Parker}, {Norton}, \&
  {Mukai}}]{ParkerEtAl2005}
{Parker}, T.~L., {Norton}, A.~J., \& {Mukai}, K. 2005, \aap, 439, 213

\bibitem[{{Patterson}(1979)}]{Patterson1979}
{Patterson}, J. 1979, \apjl, 233, L13

\bibitem[{{Patterson}(1994)}]{Patterson1994}
{Patterson}, J. 1994, \pasp, 106, 209

\bibitem[{{Patterson} {et~al.}(2002){Patterson}, {Fenton}, {Thorstensen},
  {Harvey}, {Skillman}, {Fried}, {Monard}, {O'Donoghue}, {Beshore}, {Martin},
  {Niarchos}, {Vanmunster}, {Foote}, {Bolt}, {Rea}, {Cook}, {Butterworth}, \&
  {Wood}}]{PattersonEtAl2002}
{Patterson}, J., {Fenton}, W.~H., {Thorstensen}, J.~R., {et~al.} 2002, \pasp,
  114, 1364

\bibitem[{{Patterson} {et~al.}(1978){Patterson}, {Robinson}, \&
  {Nather}}]{PattersonEtAl1978}
{Patterson}, J., {Robinson}, E.~L., \& {Nather}, R.~E. 1978, \apj, 224, 570

\bibitem[{{Pereira} {et~al.}(2007){Pereira}, {McGaha}, {Young}, \&
  {Rhoades}}]{PereiraEtAl2007}
{Pereira}, A.~J.~S., {McGaha}, J.~E., {Young}, J., \& {Rhoades}, H. 2007,
  International Astronomical Union Circular, 8895, 1

\bibitem[{{Poole} {et~al.}(2008){Poole}, {Breeveld}, {Page}, {Landsman},
  {Holland}, {Roming}, {Kuin}, {Brown}, {Gronwall}, {Hunsberger}, {Koch},
  {Mason}, {Schady}, {vanden Berk}, {Blustin}, {Boyd}, {Broos}, {Carter},
  {Chester}, {Cucchiara}, {Hancock}, {Huckle}, {Immler}, {Ivanushkina},
  {Kennedy}, {Marshall}, {Morgan}, {Pandey}, {de Pasquale}, {Smith}, \&
  {Still}}]{PooleEtAl2008}
{Poole}, T.~S., {Breeveld}, A.~A., {Page}, M.~J., {et~al.} 2008, \mnras, 383,
  627

\bibitem[{{Pretorius} \& {Knigge}(2008)}]{PretoriusKnigge2008}
{Pretorius}, M.~L. \& {Knigge}, C. 2008, \mnras, 385, 1485

\bibitem[{{Pretorius} \& {Mukai}(2014)}]{PretoriusMukai2014}
{Pretorius}, M.~L. \& {Mukai}, K. 2014, \mnras, 442, 2580

\bibitem[{{Retter} {et~al.}(1998){Retter}, {Leibowitz}, \&
  {Kovo-Kariti}}]{RetterEtAl1998}
{Retter}, A., {Leibowitz}, E.~M., \& {Kovo-Kariti}, O. 1998, \mnras, 293, 145

\bibitem[{{Revnivtsev} {et~al.}(2009){Revnivtsev}, {Sazonov}, {Churazov},
  {Forman}, {Vikhlinin}, \& {Sunyaev}}]{RevnivtsevEtAl2009}
{Revnivtsev}, M., {Sazonov}, S., {Churazov}, E., {et~al.} 2009, \nat, 458, 1142

\bibitem[{{Robinson} \& {Nather}(1983)}]{RobinsonNather1983}
{Robinson}, E.~L. \& {Nather}, R.~E. 1983, \apj, 273, 255

\bibitem[{{Rodr{\'{\i}}guez-Gil} \&
  {Mart{\'{\i}}nez-Pais}(2002)}]{Rodriguez-GilMartinez-Pais2002}
{Rodr{\'{\i}}guez-Gil}, P. \& {Mart{\'{\i}}nez-Pais}, I.~G. 2002, \mnras, 337,
  209

\bibitem[{{Rodr{\'{\i}}guez-Gil} {et~al.}(2009){Rodr{\'{\i}}guez-Gil},
  {Mart{\'{\i}}nez-Pais}, \& {de la Cruz
  Rodr{\'{\i}}guez}}]{RodriguezGilEtAl2009}
{Rodr{\'{\i}}guez-Gil}, P., {Mart{\'{\i}}nez-Pais}, I.~G., \& {de la Cruz
  Rodr{\'{\i}}guez}, J. 2009, \mnras, 395, 973

\bibitem[{{Rodr{\'{\i}}guez-Gil} \& {Torres}(2005)}]{Rodriguez-GilTorres2005}
{Rodr{\'{\i}}guez-Gil}, P. \& {Torres}, M.~A.~P. 2005, \aap, 431, 289

\bibitem[{{Schwarz} {et~al.}(2011){Schwarz}, {Ness}, {Osborne}, {Page},
  {Evans}, {Beardmore}, {Walter}, {Helton}, {Woodward}, {Bode}, {Starrfield},
  \& {Drake}}]{SchwarzEtAl2011}
{Schwarz}, G.~J., {Ness}, J.-U., {Osborne}, J.~P., {et~al.} 2011, \apjs, 197,
  31

\bibitem[{{Schwarzenberg-Czerny}(1989)}]{Schwarzenberg-Czerny1989}
{Schwarzenberg-Czerny}, A. 1989, \mnras, 241, 153

\bibitem[{{Silber} {et~al.}(1996){Silber}, {Anderson}, {Margon}, \&
  {Downes}}]{SilberEtAl1996}
{Silber}, A.~D., {Anderson}, S.~F., {Margon}, B., \& {Downes}, R.~A. 1996,
  \apj, 462, 428

\bibitem[{{Skiff}(1995)}]{Skiff1995}
{Skiff}, B. 1995, \iaucirc, 6181

\bibitem[{{Skrutskie} {et~al.}(2006){Skrutskie}, {Cutri}, {Stiening},
  {Weinberg}, {Schneider}, {Carpenter}, {Beichman}, {Capps}, {Chester},
  {Elias}, {Huchra}, {Liebert}, {Lonsdale}, {Monet}, {Price}, {Seitzer},
  {Jarrett}, {Kirkpatrick}, {Gizis}, {Howard}, {Evans}, {Fowler}, {Fullmer},
  {Hurt}, {Light}, {Kopan}, {Marsh}, {McCallon}, {Tam}, {Van Dyk}, \&
  {Wheelock}}]{SkrutskieEtAl2006}
{Skrutskie}, M.~F., {Cutri}, R.~M., {Stiening}, R., {et~al.} 2006, \aj, 131,
  1163

\bibitem[{{Southworth} {et~al.}(2008){Southworth}, {G{\"a}nsicke}, {Marsh},
  {Torres}, {Steeghs}, {Hakala}, {Copperwheat}, {Aungwerojwit}, \&
  {Mukadam}}]{SouthworthEtAl2008}
{Southworth}, J., {G{\"a}nsicke}, B.~T., {Marsh}, T.~R., {et~al.} 2008, \mnras,
  391, 591

\bibitem[{{Staude} {et~al.}(2006){Staude}, {Schwope}, {Schwarz}, {Vogel}, \&
  {Krumpe}}]{StaudeEtAl2006}
{Staude}, A., {Schwope}, A., {Schwarz}, R., {Vogel}, J., \& {Krumpe}, M. 2006,
  in ESA Special Publication, Vol. 604, The X-ray Universe 2005, ed.
  A.~{Wilson}, 307

\bibitem[{{Stratton}(1934)}]{Stratton1934}
{Stratton}, F.~J.~M. 1934, \nat, 134, 974

\bibitem[{{Str{\"u}der} {et~al.}(2001){Str{\"u}der}, {Briel}, {Dennerl},
  {Hartmann}, {Kendziorra}, {Meidinger}, {Pfeffermann}, {Reppin}, {Aschenbach},
  {Bornemann}, {Br{\"a}uninger}, {Burkert}, {Elender}, {Freyberg}, {Haberl},
  {Hartner}, {Heuschmann}, {Hippmann}, {Kastelic}, {Kemmer}, {Kettenring},
  {Kink}, {Krause}, {M{\"u}ller}, {Oppitz}, {Pietsch}, {Popp}, {Predehl},
  {Read}, {Stephan}, {St{\"o}tter}, {Tr{\"u}mper}, {Holl}, {Kemmer}, {Soltau},
  {St{\"o}tter}, {Weber}, {Weichert}, {von Zanthier}, {Carathanassis}, {Lutz},
  {Richter}, {Solc}, {B{\"o}ttcher}, {Kuster}, {Staubert}, {Abbey}, {Holland},
  {Turner}, {Balasini}, {Bignami}, {La Palombara}, {Villa}, {Buttler},
  {Gianini}, {Lain{\'e}}, {Lumb}, \& {Dhez}}]{StruderEtAl2001}
{Str{\"u}der}, L., {Briel}, U., {Dennerl}, K., {et~al.} 2001, \aap, 365, L18

\bibitem[{{Szkody} {et~al.}(2003){Szkody}, {Fraser}, {Silvestri}, {Henden},
  {Anderson}, {Frith}, {Lawton}, {Owens}, {Raymond}, {Schmidt}, {Wolfe},
  {Bochanski}, {Covey}, {Harris}, {Hawley}, {Knapp}, {Margon}, {Voges},
  {Walkowicz}, {Brinkmann}, \& {Lamb}}]{SzkodyEtAl2003}
{Szkody}, P., {Fraser}, O., {Silvestri}, N., {et~al.} 2003, \aj, 126, 1499

\bibitem[{{Thorstensen} \& {Taylor}(2000)}]{ThorstensenTaylor2000}
{Thorstensen}, J.~R. \& {Taylor}, C.~J. 2000, \mnras, 312, 629

\bibitem[{{Tody}(1993)}]{Tody1993}
{Tody}, D. 1993, in Astronomical Society of the Pacific Conference Series,
  Vol.~52, Astronomical Data Analysis Software and Systems II, ed. R.~J.
  {Hanisch}, R.~J.~V. {Brissenden}, \& J.~{Barnes}, 173

\bibitem[{{Tonry} {et~al.}(2012){Tonry}, {Stubbs}, {Lykke}, {Doherty},
  {Shivvers}, {Burgett}, {Chambers}, {Hodapp}, {Kaiser}, {Kudritzki},
  {Magnier}, {Morgan}, {Price}, \& {Wainscoat}}]{TonryEtAl2012}
{Tonry}, J.~L., {Stubbs}, C.~W., {Lykke}, K.~R., {et~al.} 2012, \apj, 750, 99

\bibitem[{{Turner} {et~al.}(2001){Turner}, {Abbey}, {Arnaud}, {Balasini},
  {Barbera}, {Belsole}, {Bennie}, {Bernard}, {Bignami}, {Boer}, {Briel},
  {Butler}, {Cara}, {Chabaud}, {Cole}, {Collura}, {Conte}, {Cros}, {Denby},
  {Dhez}, {Di Coco}, {Dowson}, {Ferrando}, {Ghizzardi}, {Gianotti}, {Goodall},
  {Gretton}, {Griffiths}, {Hainaut}, {Hochedez}, {Holland}, {Jourdain},
  {Kendziorra}, {Lagostina}, {Laine}, {La Palombara}, {Lortholary}, {Lumb},
  {Marty}, {Molendi}, {Pigot}, {Poindron}, {Pounds}, {Reeves}, {Reppin},
  {Rothenflug}, {Salvetat}, {Sauvageot}, {Schmitt}, {Sembay}, {Short},
  {Spragg}, {Stephen}, {Str{\"u}der}, {Tiengo}, {Trifoglio}, {Tr{\"u}mper},
  {Vercellone}, {Vigroux}, {Villa}, {Ward}, {Whitehead}, \&
  {Zonca}}]{TurnerEtAl2001}
{Turner}, M.~J.~L., {Abbey}, A., {Arnaud}, M., {et~al.} 2001, \aap, 365, L27

\bibitem[{{Vaytet} {et~al.}(2007){Vaytet}, {O'Brien}, \&
  {Rushton}}]{VaytetEtAl2007}
{Vaytet}, N.~M.~H., {O'Brien}, T.~J., \& {Rushton}, A.~P. 2007, \mnras, 380,
  175

\bibitem[{{Voges} {et~al.}(1999){Voges}, {Aschenbach}, {Boller},
  {Br{\"a}uninger}, {Briel}, {Burkert}, {Dennerl}, {Englhauser}, {Gruber},
  {Haberl}, {Hartner}, {Hasinger}, {K{\"u}rster}, {Pfeffermann}, {Pietsch},
  {Predehl}, {Rosso}, {Schmitt}, {Tr{\"u}mper}, \&
  {Zimmermann}}]{VogesEtAl1999}
{Voges}, W., {Aschenbach}, B., {Boller}, T., {et~al.} 1999, \aap, 349, 389

\bibitem[{{Walker}(1956)}]{Walker1956}
{Walker}, M.~F. 1956, \apj, 123, 68

\bibitem[{{Warner} \& {Woudt}(2009)}]{WarnerWoudt2009}
{Warner}, B. \& {Woudt}, P.~A. 2009, \mnras, 397, 979

\bibitem[{{Warwick} {et~al.}(2014){Warwick}, {Byckling}, \&
  {P{\'e}rez-Ram{\'{\i}}rez}}]{WarwickEtAl2014}
{Warwick}, R.~S., {Byckling}, K., \& {P{\'e}rez-Ram{\'{\i}}rez}, D. 2014,
  \mnras, 438, 2967

\bibitem[{{Wolf} {et~al.}(2018){Wolf}, {Onken}, {Luvaul}, {Schmidt}, {Bessell},
  {Chang}, {Da Costa}, {Mackey}, {Martin-Jones}, {Murphy}, {Preston}, {Scalzo},
  {Shao}, {Smillie}, {Tisserand}, {White}, \& {Yuan}}]{WolfEtAl2018}
{Wolf}, C., {Onken}, C.~A., {Luvaul}, L.~C., {et~al.} 2018, \pasa, 35, e010

\bibitem[{{Worpel} {et~al.}(2018){Worpel}, {Schwope}, {Traulsen}, {Mukai}, \&
  {Ok}}]{WorpelEtAl2018}
{Worpel}, H., {Schwope}, A.~D., {Traulsen}, I., {Mukai}, K., \& {Ok}, S. 2018,
  \aap, 617, A52

\bibitem[{{Worrall} {et~al.}(1982){Worrall}, {Marshall}, {Boldt}, \&
  {Swank}}]{WorrallEtAl1982}
{Worrall}, D.~M., {Marshall}, F.~E., {Boldt}, E.~A., \& {Swank}, J.~H. 1982,
  \apj, 255, 111

\bibitem[{{Woudt} {et~al.}(2012){Woudt}, {Warner}, {Gulbis}, {Coppejans},
  {Hambsch}, {Beardmore}, {Evans}, {Osborne}, {Page}, {Wynn}, \& {van der
  Heyden}}]{WoudtEtAl2012}
{Woudt}, P.~A., {Warner}, B., {Gulbis}, A., {et~al.} 2012, \mnras, 427, 1004

\bibitem[{{Woudt} {et~al.}(2004){Woudt}, {Warner}, \&
  {Pretorius}}]{WoudtEtAl2004}
{Woudt}, P.~A., {Warner}, B., \& {Pretorius}, M.~L. 2004, \mnras, 351, 1015

\bibitem[{{Woudt} {et~al.}(2005){Woudt}, {Warner}, \& {Spark}}]{WoudtEtAl2005}
{Woudt}, P.~A., {Warner}, B., \& {Spark}, M. 2005, \mnras, 364, 107

\bibitem[{{Wright} {et~al.}(2010){Wright}, {Eisenhardt}, {Mainzer}, {Ressler},
  {Cutri}, {Jarrett}, {Kirkpatrick}, {Padgett}, {McMillan}, {Skrutskie},
  {Stanford}, {Cohen}, {Walker}, {Mather}, {Leisawitz}, {Gautier}, {McLean},
  {Benford}, {Lonsdale}, {Blain}, {Mendez}, {Irace}, {Duval}, {Liu}, {Royer},
  {Heinrichsen}, {Howard}, {Shannon}, {Kendall}, {Walsh}, {Larsen}, {Cardon},
  {Schick}, {Schwalm}, {Abid}, {Fabinsky}, {Naes}, \& {Tsai}}]{WrightEtAl2010}
{Wright}, E.~L., {Eisenhardt}, P.~R.~M., {Mainzer}, A.~K., {et~al.} 2010, \aj,
  140, 1868

\bibitem[{{Zhang} {et~al.}(1995){Zhang}, {Robinson}, {Stiening}, \&
  {Horne}}]{ZhangEtAl1995}
{Zhang}, E., {Robinson}, E.~L., {Stiening}, R.~F., \& {Horne}, K. 1995, \apj,
  454, 447

\end{thebibliography}

\end{document}